\documentclass[twocolumn,floatfix]{aastex63}

\usepackage{graphicx} 
\graphicspath{{figures/}} 

\usepackage{dcolumn} 
\usepackage{bm} 
\usepackage{hyperref} 
\usepackage{float} 
\usepackage{booktabs} 
\usepackage{enumitem} 

\usepackage{amsmath} 
\usepackage{amssymb} 
\usepackage{latexsym}

\usepackage{cleveref}

\usepackage[english]{babel}
\usepackage[expansion=all,babel]{microtype}

\newcommand{\Msun}{\mathrm{M}_\odot}

\newcommand{\Mmax}{M_\mathrm{max}}

\newcommand{\rhob}{\rho_\mathrm{b}}
\newcommand{\rhoc}{\rho_\mathrm{c}}
\newcommand{\Ts}{T_\mathrm{eff}}
\newcommand{\Tb}{T_\mathrm{b}}
\newcommand{\gs}{g_\mathrm{s}}

\newcommand{\dd}{\mathrm{d}}
\newcommand{\sigmaSB}{\sigma_\mathrm{SB}}
\newcommand{\gcc}{\mathrm{g~cm^{-3}}}
\newcommand{\slfrac}[2]{\left.#1\middle/#2\right.}
\newcommand{\e}{\mathrm{e}}

\makeatletter
	\newcommand{\vast}{\bBigg@{3.4}}
\makeatother

\newcommand{\Eq}[1]{Eq.~(\ref{#1})}
\newcommand{\Fig}[1]{Fig.~\ref{#1}}

\newcommand{\Sec}[1]{Sect.~\ref{#1}}
\newcommand{\App}[1]{Appendix~\ref{#1}}

\newcommand{\rev}[1]{\textcolor{black}{#1}}

\submitjournal{ApJ}

\shorttitle{Evolution of Neo Neutron Stars. I.} 
\shortauthors{Beznogov, Page, and Ramirez-Ruiz}

\begin{document}

\title{THERMAL EVOLUTION OF NEO-NEUTRON STARS. I: ENVELOPES, EDDINGTON LUMINOSITY PHASE AND IMPLICATIONS FOR GW170817}
\author[0000-0002-7326-7270]{Mikhail V. Beznogov}
\affiliation{Instituto de Astronom\'ia, Universidad Nacional Aut\'onoma de M\'exico, Ciudad de  M\'exico, 04510, Mexico}
\email{mikhail@astro.unam.mx}

\author[0000-0003-2498-4326]{Dany Page}
\affiliation{Instituto de Astronom\'ia, Universidad Nacional Aut\'onoma de M\'exico, Ciudad de  M\'exico, 04510, Mexico}
\email{page@astro.unam.mx}

\author[0000-0003-2558-3102]{Enrico Ramirez-Ruiz}
\affiliation{Department of Astronomy and Astrophysics, University of California, Santa Cruz, CA 95064, USA}
\affiliation{Niels Bohr Institute, University of Copenhagen, Blegdamsvej 17, 2100 Copenhagen, Denmark}
\email{enrico@ucolick.org}

\date{\today}
             
\begin{abstract}
A neo-neutron star is a hot neutron star that has just become transparent to neutrinos.
In a core collapse supernova or accretion induced collapse of a white dwarf the neo-neutron star phase directly follows the proto-neutron star phase,
about 30 to 60 seconds after the initial collapse.
It will also be present in a binary neutron star merger in the case the ``born-again'' hot massive compact star does not immediately collapse into a black hole.
Eddington or even super-Eddington luminosities are present for some time. 
A neo-neutron star produced in a core collapse supernova is not directly observable but the one produced by a binary merger, 
likely associated with an off-axis short gamma-ray burst, may be observable for some time 
as well as when produced in the accretion induced collapse of a white dwarf.
We present a first step in the study of this neo-neutron star phase in a spherically symmetric configuration, thus neglecting fast rotation, and also
neglecting the effect of strong magnetic fields.
We put particular emphasis on determining how long the star can sustain a near-Eddington luminosity
and also show the importance of positrons and contraction energy during neo-neutron star phase.
We finally discuss the observational prospects for neutron star mergers triggered by LIGO and for accretion-induced collapse transients.
\end{abstract}

\section{Introduction}
\label{sec:Intro}

Neutron stars are by far the most intriguing objects in the Universe. 
They are superdense, can be superfast rotators, may have superstrong magnetic fields, and are surrounded by the strongest gravitational fields \citep[see, e.g.][]{HPY07}. 
They are born in core collapse supernova events \citep{1934PNAS...20..254B} or in
accretion induced collapse of white dwarves \citep{1976A&A....46..229C} and start their life as proto-neutron stars \citep{Burrows:1986aa}.
Moreover, a hot born-again massive neutron star may also be produced in the merging of a binary neutron star system and survive as such \citep{Kluzniak:1998uq},
or collapse into a black hole.
During the first hot phase, neutrinos are copiously produced but are trapped in the stellar interior and only escape by slowly diffusing outward.
This early evolution, lasting less than a minute, has been extensively studied theoretically, in large part because a Galactic core-collapse supernova would allow us to
follow it observationally through the detection of the emitted neutrinos.
The subsequent phase, which we will call the {\em neo-neutron star} phase,
from an age of a minute after the birth/re-birth to a few hours/days, has, however, never been carefully considered. 
Later phases have been the object of numerous studies (see, e.g., \citealt*{YP04,PGW06}).

After the supernova, it may take decades till the ejecta become transparent to electromagnetic radiation from the central object \citep{1970ApJ...162..737B}.
In the case of the supernova SN 1987A it is only recently, after more than thirty years, that credible evidence of the presence of a compact
object has been found \citep{Cigan:2019aa}.
The youngest observed neutron star is the compact object in the center of the Cassiopeia A supernova remnant \citep{Tananbaum:1999aa}, 
with an age of about 340 years \citep{Fesen:2006aa}.
It is thus doubtful we will have, in the near future, valuable observational data on the very early cooling history of a neutron star, and even less of a neo-neutron star.
The neo-neutron phase is, however, the phase during which the neutron star crust is formed and it is, thus, establishing the basic structure
for a large amount of neutron star phenomenology.

A complementary set-up is provided by binary neutron star mergers \citep{2007NJPh....9...17L,2012LRR....15....8F}.
Although it is often considered that the outcome of such event would be the formation of a low-mass black hole \citep{1989Natur.340..126E,2011ApJ...732L...6R,2014ApJ...788L...8M},
there is a possibility that the merged object survives as a massive neutron star \citep{1992Natur.357..472U,Kluzniak:1998uq,2008MNRAS.385.1455M}.
In such a scenario we would have a ``born-again'' neutron star, with trapped neutrinos because of its high temperature,
followed by a massive neo-neutron star.
This possibility is real only if the high density equation of state (EOS) is stiff enough to have a high maximum mass, $\Mmax$.
The maximum mass of a cold slowly- or non-rotating neutron star is at least $2\, \Msun$, from the masses of the pulsars
PSR J1614-2230 \citep{Demorest_etal10} and PSR J0348+0432 \citep{Antoniadis_etal13}
and possibly higher than $2.3\, \Msun$ from the upper value of the mass of PSR J0740+6620 \citep{2019arXiv190406759C}.
Analyses of the GW170817/GRB170817A gravitational wave/gamma-ray burst event have also provided new constraints on $\Mmax$
based on the delayed collapse of the merged object into a black hole.
 \citet{Margalit:2017aa} obtain $\Mmax \leq 2.17\, \Msun$ (90\%) and 
\citet{Rezzolla:2018aa} find $2\, \Msun \leq \Mmax \leq 2.3\, \Msun$
while the more detailed study of \citet{Shibata:2019aa} conclude that $ \Mmax \leq 2.3\, \Msun$.
If such is the case only mergers of binaries containing low mass neutron stars could produce a stable merged object so that
our neo-neutron star description would be of interest.

As a first step, in the present paper, we consider the evolution of the outer layer of the neo-neutron star, its envelope,
just after the formation of nuclei when the surface temperature is high enough to be of the order of the Eddington luminosity,
i.e., of the order of $10^{38}$~erg~s$^{-1}$.
An important question we tackle is the duration of a possible Eddington or super-Eddington phase, and then consider the subsequent evolution.
Based on previous studies of proto-neutron star we explore the impact of different possible initial temperature/luminosity profiles in the envelope
on the Eddington phase. 

This paper is structured as follows. In \Sec{sec:MainEq} and \ref{sec:Boundary} we setup the problem and present our results on the low-density regime of an Eddington envelope.
In \Sec {sec:NeoNS} we define what a neo-netron star phase is and
contrast the method for study of long term cooling of isolated neutron stars
versus the neo-neutron star case.
Our resulted are described in \Sec{sec:Results} and their observational relevance is discussed in \Sec{sec:obs}.
A summary and conclusions are presented in \Sec{sec:Concl}. 
Finally in \App{sec:NeoNS:Phys} and \ref{app:NCv} we describe the 
physical properties of hot neutron star envelopes and our numerical scheme 
is detailed in \App{sec:Solver}.

\section{Thermal Evolution Equations}
\label{sec:MainEq}

We consider a spherically symmetric problem, neglecting the effects of rotation and magnetic fields.
Since the structure of the outer layers of our stars will expand or contract 
we employ the enclosed baryon number $a$ as a (Lagrange) radial variable
instead of the circumferential radius $r$.
The full set of general relativistic structure and mechanical evolution equations 
can be found, e.g., in \cite{Potekhin:2018jb}.
The thermal evolution equations we will solve are:
\begin{align}
	&\widetilde{L} = -K \left(4\pi r^2 \right) ^2 n \e^\phi \frac{\partial \widetilde{T}}{\partial a},
\label{Eq:Thermal_T}   \\
	\begin{split}
		&\e^\phi \frac{\partial (\widetilde{T} \e^{-\phi})}{\partial t} = 
		-\frac{1}{C_\mathrm{V}}(\widetilde{Q}_\mathrm{L} + \widetilde{Q}_\nu  + \widetilde{Q}_\mathrm{V})
	\end{split}	
\label{Eq:Thermal_L} 
\end{align}
where $\widetilde{L} = L \e^{2\phi}$ and $\widetilde{T} = T \e^{\phi}$ are red-shifted luminosity and temperature, respectively,
$\e^{2\phi}$ being the time component of the metric.
$C_\mathrm{V}$ is the heat capacity and $K$ the thermal conductivity.
The energy sources/sinks are
\begin{equation}
\widetilde{Q}_\mathrm{L} \equiv n \frac{\partial \widetilde{L}}{\partial a}
\label{Eq:QL}
\end{equation}
which gives the heat loss/injection from the luminosity gradient, 
$\widetilde{Q}_\nu = \e^{2\phi} Q_\nu$ which gives the neutrino energy loss and
\begin{equation}
\widetilde{Q}_\mathrm{V} \equiv - \widetilde{T} \left.  \left( \frac{\partial P}{\partial T}\right)\right|_n \frac{\partial \ln n}{\partial t}
\label{Eq:QV}
\end{equation}
that will be called the \emph{contraction energy}.
This last term comes from the ``$P \, \dd V$'' work and the volume dependent part of the internal energy and 
gives the gravitational and internal energy release owing to the contraction of the star during its cooling.

We follow the evolution of the star with a sequence of models in 
hydrostatic equilibrium, i.e., we neglect acceleration compared to gravity.
We have explicitly checked that our results are consistent with this 
approximation.

The microphysics we employ is standard and we describe it in 
\App{sec:NeoNS:Phys} and \ref{app:NCv}.


\section{Outer Boundary and Envelope}
\label{sec:Boundary}

Boundary conditions at the center, where $a=0$, are obvious:
\begin{align}
	\widetilde{L}(0) = 0, \quad r(0) = 0, \quad m(0) = 0,
\label{Eq:Boundary_center}
\end{align}
while $P(a=0) \equiv P_\mathrm{c}$ is an arbitrary parameter that will determine the mass of the star.
Boundary conditions at the surface are more delicate and ``surface'' must be properly defined.
The simplest and naive condition is the ``zero condition'': $P_\mathrm{s} = \rho_\mathrm{s} = T_\mathrm{s}=0$, and one takes $R \equiv r_\mathrm{s}$ and $M \equiv m_\mathrm{s}$.
Subscript ``s'' refers to the quantities at the surface.
However, this is too naive since $P$, $\rho$, and $T$, likely never really reach zero and, more likely, there is a smooth transition from the stellar interior to
the surrounding magnetosphere (or the interstellar medium in the case of a non magnetized star).
It is more appropriate, when studying the thermal evolution of the star, to define its surface as located at the photosphere, i.e.,
the layer where the outflowing thermal radiation is produced.
We adopt the commonly used {\em Eddington, or photospheric, condition} (see, e.g., \citealt*{HKT04}) in which detailed radiative transfer 
(where the energy dependence of the opacity is wholly taken into account) is replaced by a diffusion approximation
(where the energy dependent opacity is replaced by its Rosseland mean),
i.e., the same equation (\ref{Eq:Thermal_T}), and the photosphere is defined as the layer where the optical depth is $2/3$.
This lead to the conditions
\begin{align}
		L_\mathrm{s} = 4 \pi \sigmaSB R^2 T_\mathrm{s}^4
	\label{Eq:Boundary_surfL}
\end{align}
and
\begin{align}
		P_\mathrm{s} = \frac{2}{3}\frac{\gs}{\kappa_\mathrm{s}} \left(1+ \frac{L_\mathrm{s}}{L_\mathrm{Edd}}  \right)
	\label{Eq:Boundary_surfP}
\end{align}
where 
\begin{equation}
\gs = \e^\lambda \, GM/R^2
\label{Eq:gs}
\end{equation}
is the free-fall acceleration at the surface, 
$\e^{2\lambda}$ being the radial component of the metric,
$\sigmaSB$ the Stephan-Boltzmann constant, $\kappa_\mathrm{s}$
the Rosseland mean opacity, at the surface, and 
\begin{equation}
L_\mathrm{Edd}(R) = \frac{4\pi c \,GM \e^\lambda}{\kappa_\mathrm{s}} = \frac{4\pi R^2 \, c \, \gs}{\kappa_\mathrm{s}}
\label{Eq:LEdd}
\end{equation}
the Eddington luminosity at the stellar surface.

These are then complemented by the obvious relations that define the mass $M$ and radius $R$ of the star
\begin{align}
		M = m_\mathrm{s}  \quad \text{and} \quad R = r_\mathrm{s},
\end{align}
and the continuity of the metric coefficient with the external Schwarzschild solution		
\begin{align}		
		\e^{\phi(R)} = \e^{-\lambda(R)} = \sqrt{1-2GM/Rc^2} \quad .
\end{align}

It is numerically inconvenient to directly apply the outer boundary conditions of \Eq{Eq:Boundary_surfL} and (\ref{Eq:Boundary_surfP}) 
and we rather apply the standard scheme of separating out an {\em envelope} \citep*{Gudmundsson:1982yq} as described below.

In these outer layers 
the equation of hydrostatic equilibrium is simply
\begin{equation}
\frac{\dd P}{\dd l} = - \gs \rho
\label{Eq:hydro}
\end{equation}
when written in terms of the proper radial length $l$ which is defined through $\dd l \equiv \e^{-\lambda(r)} \dd r$.
From a given layer at $\rho_\mathrm{a}$ and  $P_\mathrm{a}$, hydrostatic equilibrium can be integrated outward giving
the well known classical result
\begin{equation}
P_\mathrm{a} =  y_\mathrm{a} \, g_\mathrm{a} \quad \mathrm{with} \quad  y_\mathrm{a} \equiv \int_\mathrm{a}^\infty \rho(r) \dd l 
\label{Eq:column}
\end{equation}
being the proper column density of matter above point ``a'' and
where it has been assumed that the upper layers are sufficiently concentrated that $g$ can be considered constant.
In practice we can replace $g_\mathrm{a}$ by $\gs$.

\subsection{High Luminosity Envelopes}
\label{sec:Env}

The envelope is defined as the outer layers, from the surface down to a bottom layer at some pressure $P_\mathrm{b}$ or, equivalently, density $\rhob$,
in which the EOS is temperature dependent and thus require a special treatment compared to the highly degenerate interior.
For our present purpose we extended the previous models of \citet*{BPY16} to higher temperatures by adding radiation pressure to the EOS 
of fully ionized plasma of \citet{PC10}\,\footnote{The corresponding Fortran code is available at \url{http://www.ioffe.ru/astro/EIP/}}
and by adding the $L_\mathrm{s}/L_\mathrm{Edd}$ term to the surface condition as in \Eq{Eq:Boundary_surfP}.
As we justify below in \Sec{sec:NeoNS}, we restrict ourselves to envelopes made of pure iron.

\begin{figure}
	\includegraphics[height=8.0cm,keepaspectratio=true,clip=true,trim=0.45cm 0.5cm 0.7cm 0.6cm]{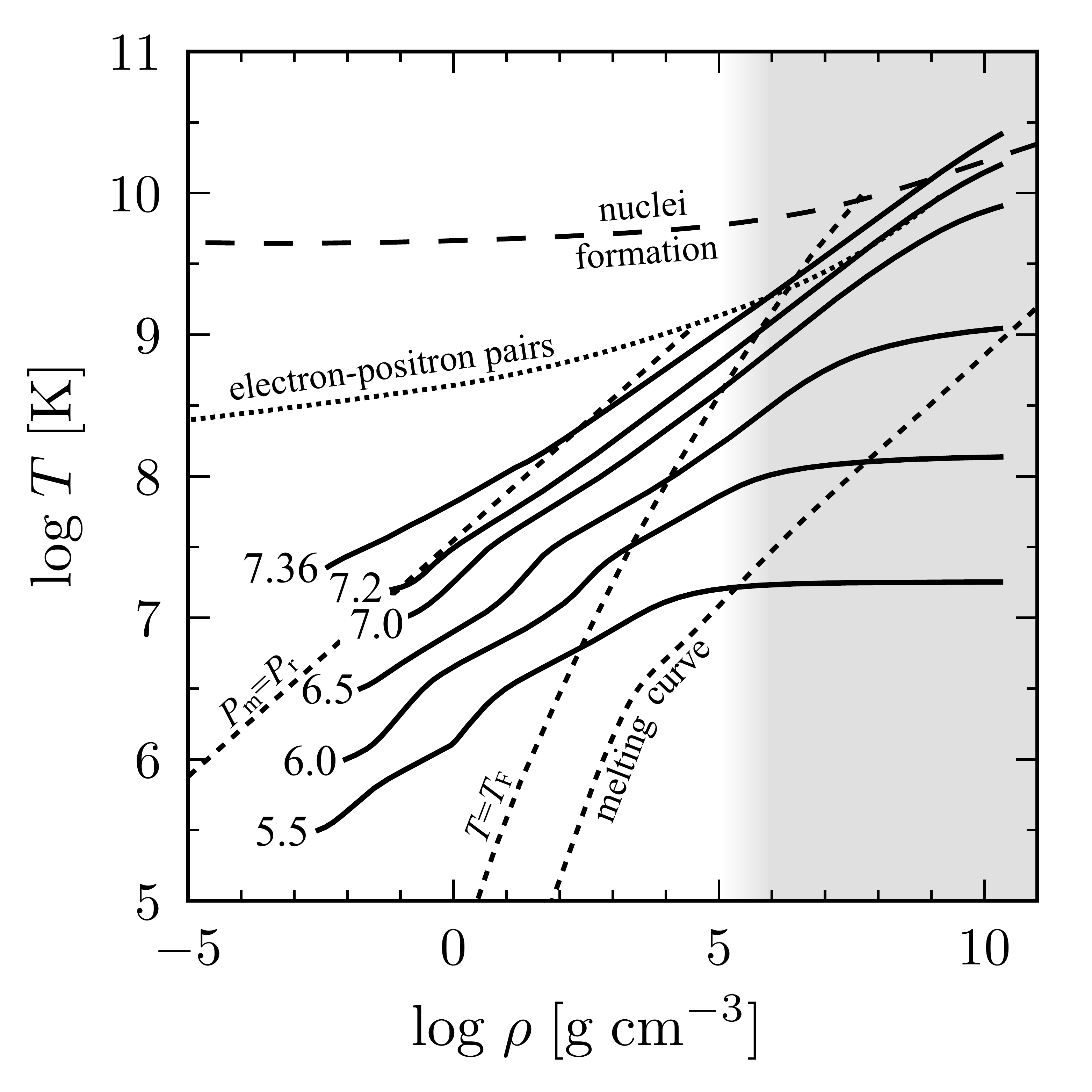}
	\caption{Envelope temperature profiles for six surfaces temperature, as labelled by $\log T_\mathrm{s}$. 
	($g_\text{s} =10^{14}$ cm s$^{-2}$ is assumed.)
	See details in the text.
	The  grey shadowed part above $10^5~\gcc$ is shown for illustration but not used in our evolutionary calculations.}
	\label{fig:Env}
\end{figure}

\Fig{fig:Env} shows six envelope temperature profiles labelled by $\log$ $T_\mathrm{s}$ and
the locations of several critical loci:
the melting curve (i.e., ions form a Coulomb crystal at densities above this curve),
the appearance of electron-positron pairs, 
the onset of electron degeneracy (labelled as ``$T=T_\mathrm{F}$''),
the transition from matter to radiation pressure dominated regimes (labelled as ``$P_\mathrm{m}=P_\mathrm{r}$''),
and the temperature below which nuclei are formed \citep*{Lattimer:1991aa}.
One can also see that the profiles with the highest temperatures
($\log$ $T_\mathrm{s}$ [K] $> 7$) cross the electron-positron pairs curve at the density about $10^5~\gcc$ and
also the nuclear formation/dissociation line when $T > 10^{10}$ K.
Neither pairs nor nuclei dissociation are included in our envelope models and these parts of the profiles should not be trusted.
However, as discussed in Sect.~\ref{sec:NeoNS}, we will locate our outer boundary density $\rhob$ at $10^5~\gcc$ and this regime of
high density inaccurately modeled envelopes will not be actually used
and is plotted here only for illustrative purpose.

Notice that in the matter dominated regime, and with opacity dominated by free-free-absorption, $\rho_\mathrm{s}$ increases with $T_\mathrm{s}$
while in the radiation dominated regime, and opacity dominated by electron scattering, the relationship is inverted.
One sees from \Fig{fig:Env} that the transition between these two regimes occurs just above $T_\mathrm{s} \sim 10^7$~K.

These envelope models provide us with a relationship between the temperature at the bottom of the envelope, $T_\mathrm{b}$, and at its surface, $T_\mathrm{s}$, the so-called ``$T_\mathrm{b} - T_\mathrm{s}$ relationship":
$T_\mathrm{b} = T_\mathrm{b}(T_\mathrm{s})$.
Since energy sources and sinks are neglected within the envelope, the luminosity at its bottom, $L_\mathrm{b}$, is equal to the surface
luminosity and thus we obtain a relationship between the two searched for solutions of \Eq{Eq:Thermal_T} and \eqref{Eq:Thermal_L}:
$L_\mathrm{b} = L_\mathrm{b}(T_\mathrm{b})$.
This allows to replace the outer boundary condition $L_\mathrm{s} = L_\mathrm{s}(T_\mathrm{s})$ of \Eq{Eq:Boundary_surfL} at $P_\mathrm{s}$
by a new one applied deeper at $P_\mathrm{b}$.

It was shown by \citet{Gudmundsson:1982yq} that in the resulting relationship  $T_\mathrm{s} = T_\mathrm{s}(T_\mathrm{b})$
the dependence on $M$ and $R$ is only through $g_\mathrm{s}$ in the form 
\begin{equation}
T_\mathrm{s}(\Tb, g_\mathrm{s, 14}) = g_\mathrm{s, 14}^{1/4} \, T_\mathrm{s}(\Tb, g_\mathrm{s, 14}=1)
\label{Eq:tbts}
\end{equation}
where $g_\mathrm{s, 14} \equiv g_\mathrm{s}/(10^{14} \, \mathrm{cm~s^{-2}})$.
We have explicitly checked that this result is still valid for our hot envelopes with a lower density inner boundary at $\rhob = 10^5~\gcc$.

We notice that the approximations that lead to the Eddington boundary condition of \Eq{Eq:Boundary_surfP} are actually self-inconsistent \citep{HKT04}
and the identification of a ``surface'' layer at temperature $T_\mathrm{s}$ has to be rather seen as a convenient {\em ansatz} for a more
realistic atmospheric boundary condition. 
It is however well-known (see, e.g., \citealt*{Kippenhahn:2012aa}) that envelope models will converge toward the ``zero condition'' and the exact definition of the
``surface'' is not important when studying the deeper layers. 
We will henceforth adopt the common notation of writing the outflowing luminosity in terms of an effective temperature $\Ts$ as
$L = 4 \pi \sigmaSB R^2 \Ts^4$ and use red-shifted quantities as $L^\infty = 4 \pi \sigmaSB R_\infty^2 \Ts^{\infty \, 4}$ with
$L^\infty \equiv \e^{2\phi} L = \widetilde{L}_\mathrm{s}$, $\Ts^\infty \equiv \e^{\phi} \Ts$ and $R_\infty \equiv \e^{-\phi} R$,
and in our case $\Ts = T_\mathrm{s}$.

\section{Neo-Neutron Stars}
\label{sec:NeoNS}

The early evolution of a newly-born, or a born-again,  neutron star can be 
divided in two separate phases:
\begin{itemize}[nosep]
	\item proto-neutron star phase, $0 \leq t \lesssim 30-60$~s.
	The star is \emph{opaque} to neutrinos, $T \gg 10^{10}$~K.
	The chemical composition of the core slowly evolves toward the zero-temperature one as neutrinos leak out
	and the star's lepton number decreases.
	\item neo-neutron star phase, $30-60~\text{s} \lesssim t \lesssim 1$ day. 
	The star becomes transparent to neutrinos, $T \ll 10^{11}$~K. The crust is being formed.
\end{itemize}
%


The standard approach used in long term cooling studies (\citealt{YP04,PGW06}) needs adjustments to study neo-neutron stars since we are now interested in very 
short term evolutions and very high temperatures.
To be able to resolve short timescales it becomes necessary to push the outer 
boundary to much lower densities and we will typically use $\rhob = 10^5~\gcc$ 
resulting in an envelope with a thermal time of the order of a second.
A direct consequence of this is that the outer layers of the interior have an EOS that becomes temperature dependent.
We thus distinguish three regions
\begin{itemize}[nosep]
	\item Outer (heat blanketing) envelope at densities $\rho_\mathrm{s} \leq \rho \leq \rhob$, treated separately in a time independent way 
	(see \Sec{sec:Env}). 
	It has \Eq{Eq:Boundary_surfL} and (\ref{Eq:Boundary_surfP}) as a surface boundary condition that defines $\rho_\mathrm{s}$, $P_\mathrm{s}$,
	and $T_\mathrm{s}$, for every given $L_\mathrm{s}$.
	\item Inner envelope in the regime $\rhob \leq \rho \leq \rhoc$ in which the EOS is still temperature dependent and where both structure
	and thermal equations have to be solved simultaneously.
	The outer envelope provides the outer boundary condition $L_\mathrm{b} = L_\mathrm{b}(T_\mathrm{b})$ for $T$ and $L$ while for
	$P$ and $\rho$ we use \Eq{Eq:column} to write $P_\mathrm{b} (t) = g_\mathrm{b}(t) y_\mathrm{b}$,
	the time dependence coming from the contraction of this inner envelope.
	With the EOS $T_\mathrm{b}(t)$ and $P_\mathrm{b}(t)$ give us $\rho_\mathrm{b}(t)$
	[and even if $P_\mathrm{b}$ is constant $\rho_\mathrm{b}$ will still change as long as $T_\mathrm{b}$ does].
	In the absence of mass loss $y_\mathrm{b}$ is constant, which is what we will assume in the present work.
	The fact that both $P$ and $\rho$, and consequently the radius $r$, change with time in the inner envelope is the reason we prefer to use the baryon number $a$,
	a conserved quantity, as radial variable. 
	\item Stellar interior at $\rho \geq \rhoc$ where the EOS is temperature independent and only the thermal equations have to be solved
	at each time step.
\end{itemize}

Models of both proto-neutron stars (see, e.g., \citealt{Burrows:1986aa}) and neutron star mergers (see, e.g., \citealt{Rosswog:2003aa})
show that the star relaxes to a temperature of a few times $10^{10}$ K in less than
a minute and so we will take as initial temperature $(2-3) \times 10^{10}$ K above $\rhoc$.
Taking $\rhoc = 10^{11}~\gcc$ as the inner boundary of the inner envelope  is sufficient to guarantee that the stellar interior EOS can be considered as temperature independent.
The microphysics we apply in the interior is the same as in long term cooling models and was described in \citet{PLPS04,PPLS11}
while the microphysics of the inner envelope is described in 
\App{sec:NeoNS:Phys}.

Numerically, solving the equations of mechanical structure and thermal 
evolution at very high temperatures where the radiation and pair pressure are 
significant 
in the inner envelope is much more challenging that in the later evolution. 
We describe in \App{sec:Solver} the details of our solver.

\section{Results}
\label{sec:Results}

\subsection{Initial Configurations}
\label{sec:Initial}


\begin{figure*}
	\includegraphics[height=8.0cm,keepaspectratio=true,clip=true,trim=0.25cm 0.5cm 0.5cm 0.6cm]{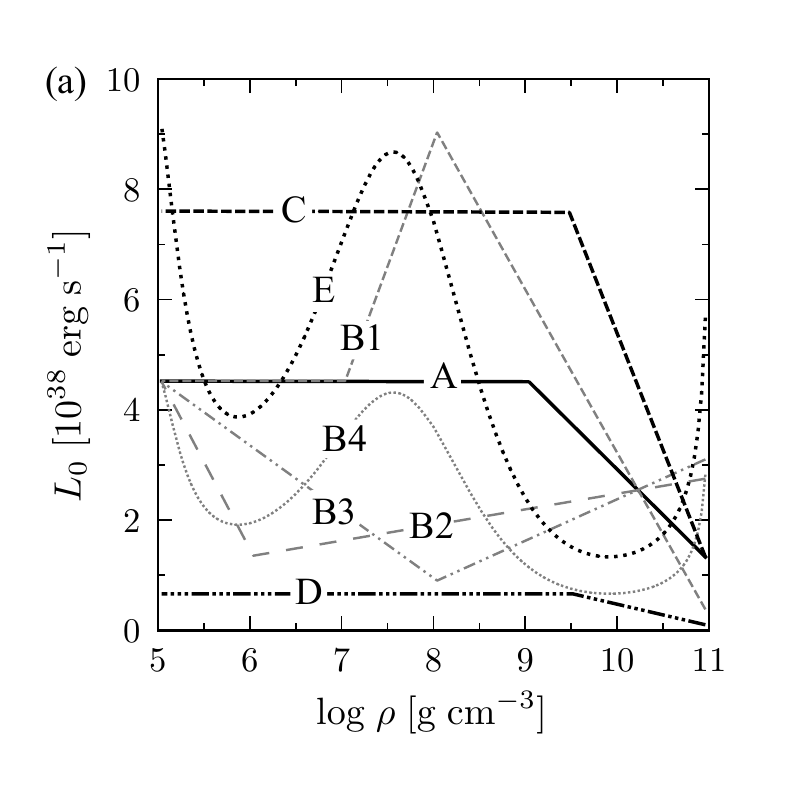}
	\hspace{0.5cm}
	\includegraphics[height=8.0cm,keepaspectratio=true,clip=true,trim=0.2cm 0.5cm 0.5cm 0.6cm]{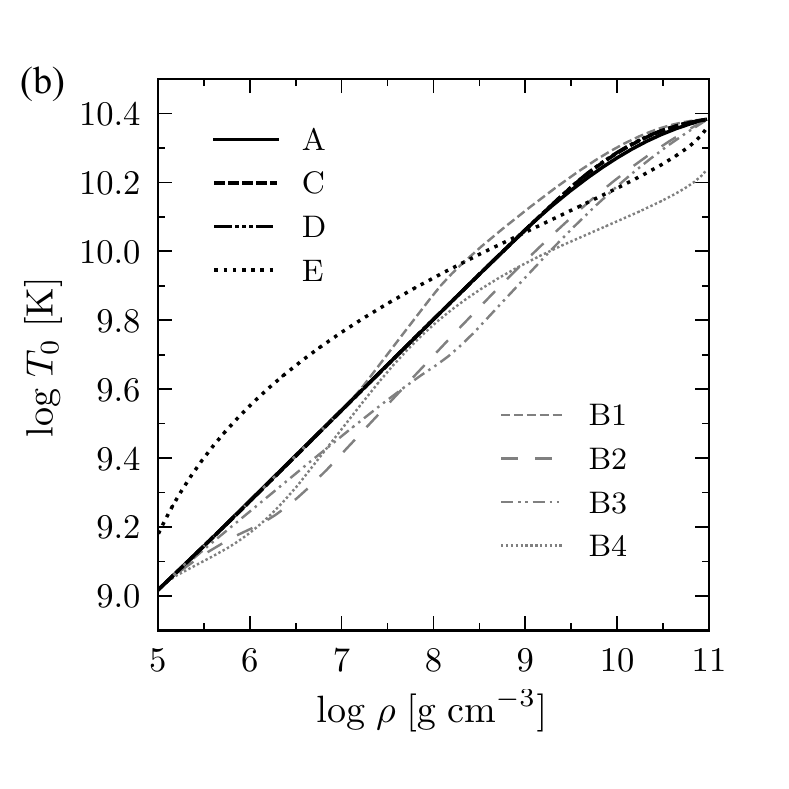}
	\caption{Panel (a): initial local luminosity profiles of all our models. 
	 Panel (b): corresponding initial local temperature profiles.
	See details in the text.}
\label{fig:Initial}
\end{figure*}

We begin our modeling once the star is transparent to neutrino, i.e. after the $\sim$ 30~s long proto-neutron star phase in the case
of a core-collapse supernova or after a similar duration after the fusion of the two stars in the case of a neutron star-neutron star merger 
(in the case the merged object survives instead of having collapsed into a black hole).
In both cases the interior temperatures are of the order of $2-3 \times 10^{10}$~K (see, e.g., \citealt{Pons:1999fk} and \citealt{Rosswog:2003aa}).
We will thus take as an initial temperature $T_0 \simeq 2.5 \times 10^{10}$~K at all densities above $\rhoc = 10^{11}~\gcc$.
At such densities this $T_0$ is just below the transition temperature where nuclei are formed in the crust 
 (\citealt{Lattimer:1991aa}; see, e.g.,  \citealt{Nakazato:2018ys} for a proto-neutron star evolution study with formation of nuclei).
We then introduce a temperature gradient in the inner envelope, from $\rhoc$ down to $\rhob = 10^5~\gcc$, our initial outer boundary point.
Numerical simulations of neither proto-neutron stars nor mergers resolve the temperature profile at low densities (outside the neutrinosphere)
and we have thus no information about this outer layer temperature gradient.
(Simulation of core-collapse supernovae do model the lower density layers but they typically follow the evolution of the system
for less than a second, see, e.g., \citealt{Janka:2012aa}.)
We want to start with a star emitting at the Eddington limit at its surface and this uniquely fixes the initial temperature $T_{\mathrm{b}, 0}$
at $\rhob$: 
the ``Eddington effective temperature'' $T_\mathrm{eff, Edd}$ is obtained, see \Eq{Eq:LEdd}, from
$L_\mathrm{Edd} = 4\pi c\, GM \e^\lambda /\kappa_\mathrm{s} = 4\pi R^2 c g_s/ \kappa_\mathrm{s} \equiv 4\pi R^2 \sigma_\mathrm{B} T_\mathrm{eff, Edd}^4$
or
\begin{equation}
T_\mathrm{eff, Edd} = \gs^{1/4} \left( \frac{c}{\sigma_\mathrm{B}\kappa_\mathrm{s}} \right)^{1/4}
\label{Eq:T_Edd}   
\end{equation} 
while envelope models relate $\Ts$ to $\Tb$ with the same $ \gs^{1/4}$ scaling, see \Eq{Eq:tbts},
implying that the $\Tb$ resulting in an Eddington luminosity is a unique temperature, $T_\mathrm{b, Edd}$, 
determined by the boundary density $\rhob$ and the chemical composition of the envelope, but independent of $M$ and $R$.
For a pure iron envelope we find that $T_\mathrm{b, Edd} = 1.07\times 10^9$~K at $\rhob = 10^5~\gcc$.

\begin{figure*}
	\includegraphics[height=8.0cm,keepaspectratio=true,clip=true,trim=0.25cm 0.5cm 0.5cm 0.6cm]{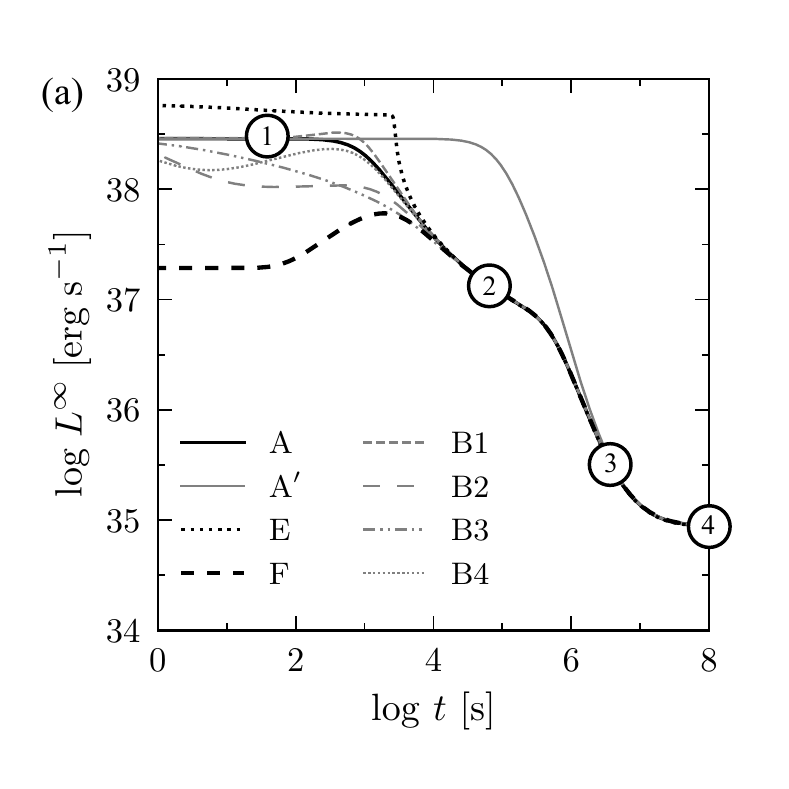}
	\hspace{0.5cm}
	\includegraphics[height=8.0cm,keepaspectratio=true,clip=true,trim=0.25cm 0.5cm 0.5cm 0.6cm]{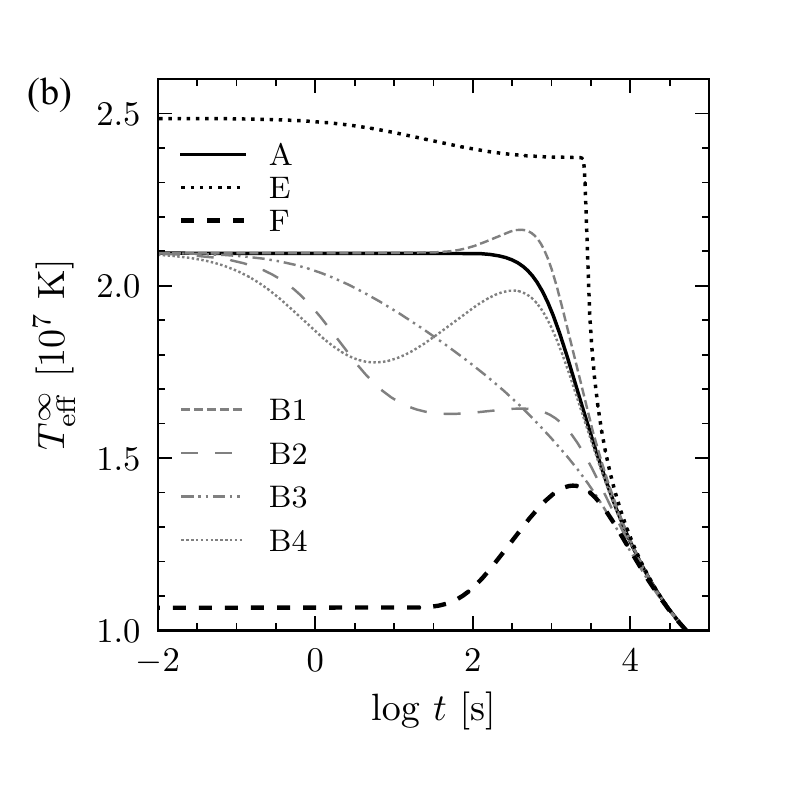}
	\caption{Cooling curves of our 1.4 $\Msun$ models A, A$^\prime$, B1, B2, B3, B4, and E, F.
	 Panel (a) shows the red-shifted luminosity $L^\infty$ and panel (b) the red-shifted effective temperature $T_\mathrm{eff}^\infty$.
	See details in the text.}
\label{fig:Cool1}
\end{figure*}

We will consider three series of stellar models, with three different surface gravities,
and implement in them different initial inner envelope luminosity or temperature profiles:
\begin{itemize}[nosep]
	\item models A, A$^\prime$, B1, B2, B3, B4, and E, F: $M = 1.4 \, \Msun$~and $R \simeq 11.6 - 11.8$ km with $g_\mathrm{s, 14} \simeq 1.6 - 1.7$.
	\item model C: $M = 2 \, \Msun$~and $R \simeq$ 11 km with $g_\mathrm{s, 14} \simeq 3.2$.
	\item models D and D$^\prime$: $M = 0.25 \, \Msun$~and $R \simeq 17 - 19$ km with $g_\mathrm{s, 14} \simeq 0.1$.
\end{itemize} 
The quoted values of the radii come from our specific choice of the core EOS from \citet*{APR98}.
The last two models D and D$^\prime$ are aimed at mimicking the effect of fast rotation where centrifugal acceleration can be seen as resulting in a small
effective surface gravity: a complete treatment of rotations would need a 2D code and our results are only intended to give a first approximation to the possible effects of fast rotation.
In model C we do not include the fast neutrino emission by the direct Urca process \citep{Boguta:1981jz,LPPH91}
acting deep in the inner core
since it has no effect on the evolution of the outer parts of the star
at early stages.

As explained in \App{sec:Solver} we find it more convenient numerically to 
define the initial luminosity  profile, $L_0(\rho)$,
in the envelope rather that defining directly $T$.
We show in \Fig{fig:Initial}a our choices: models A, B1, B2, B3, C, and D, have $L=L_\mathrm{Edd}$ at $\rhob$, with the value of $L_\mathrm{Edd}$
for their corresponding $M$, and the variation of $L$ with increasing density is constrained so that $T$ reaches 
$T_0 \simeq 2.5 \times 10^{10}$~K at $\rhoc = 10^{11}~\gcc$.
The models E is, in contradistinction, defined by the temperature profile, following \Eq{Eq:T_init}, and results in a super-Eddington $L$
at the surface.
Model B4 is obtained from the $L$ profile of model E scaled down so that its resulting surface luminosity is again the Eddington one
(but then it cannot reach $T_0$ at $\rhoc$).
In \Fig{fig:Initial}b we plot the corresponding temperature profiles.
For the reason discussed above, all models with $L(\rhob)=L_\mathrm{Edd}$ start at the same
 $T_\mathrm{b, 0} = T_\mathrm{b, Edd} =1.07\times 10^9$~K, while the super-Eddington
model E has a higher $T_\mathrm{b, 0}$.

Notice that since the total opacity $\kappa$ is much smaller in the inner envelope than at the surface, the local Eddington luminosity
$L_\mathrm{Edd}(r) = 4\pi r^2 \, c \, \gs/\kappa$ is much larger than $L_\mathrm{Edd}(R)$
and, hence, in all models the luminosity in the inner envelope is always below $L_\mathrm{Edd}(r)$.

Finally, model F represents a cold start with an initial $\Ts$ about twice lower and hence an initial surface luminosity about 15 times below $L_\mathrm{Edd}(R)$.
To avoid saturating the figure the initial $L$ and $T$ profiles of this model are not displayed in \Fig{fig:Initial}.

\subsection{Evolution of a $1.4 \, \Msun$ Star}
\label{sec:1.4Msun}

The cooling curves resulting from our initial $L$ and $T$ profiles are presented in \Fig{fig:Cool1} for our $1.4 \, \Msun$ case.
One sees that all models converge, i.e., forget their initial conditions, in about $10^4$ s (except model A$^\prime$, see below)
and this initial relaxation phase is denoted as phase ``1''.
After this, during phase ``2'', the cooling is driven by neutrino emission from the pair annihilation process and, after the 
knee\footnote{This knee can already be seen in the results of \citet{Nomoto:1987aa}, but with no interpretation provided, and in \citet{Page:1989kx}.}, 
at age $\sim 3 \times 10^5$~s, by neutrino from the plasmon decay process, the phase ``3''.
The model~A$^\prime$ has the same initial temperature and luminosity as model A but the neutrino emission by the pair annihilation process has been
arbitrarily turned off: this model confirms that pair annihilation is responsible for the evolution during phase ``2'', while
during phase ``3'' (driven by plasmon decay) model~A$^\prime$ converges toward model A. At an age of about one year the luminosity, and the surface temperature, reach a stagnation phase, ``4'': this is the ``early plateau''
already well-known in neutron star cooling studies, that will last for a few decades and corresponds to thermal relaxation of the
whole neutron star crust which will eventually reach thermal equilibrium with the core
(see \citealt{Nomoto:1987aa}, \citealt{Page:1989kx}, 
\citealt{Lattimer:1994aa}, and \citealt*{Gnedin:2001aa}).
The shift from phase ``3'' to ``4'' is due to the inner envelope temperature dropping below the plasma temperature and the
consequent exponential suppression of plasmon formation and decay: the main neutrino process available is then the very inefficient
electron-ion bremsstrahlung resulting in a significant slow-down of the cooling.

\begin{figure}
\includegraphics[height=9.0cm,keepaspectratio=true,clip=true,trim=0.0cm 0.5cm 0.7cm 0.6cm]{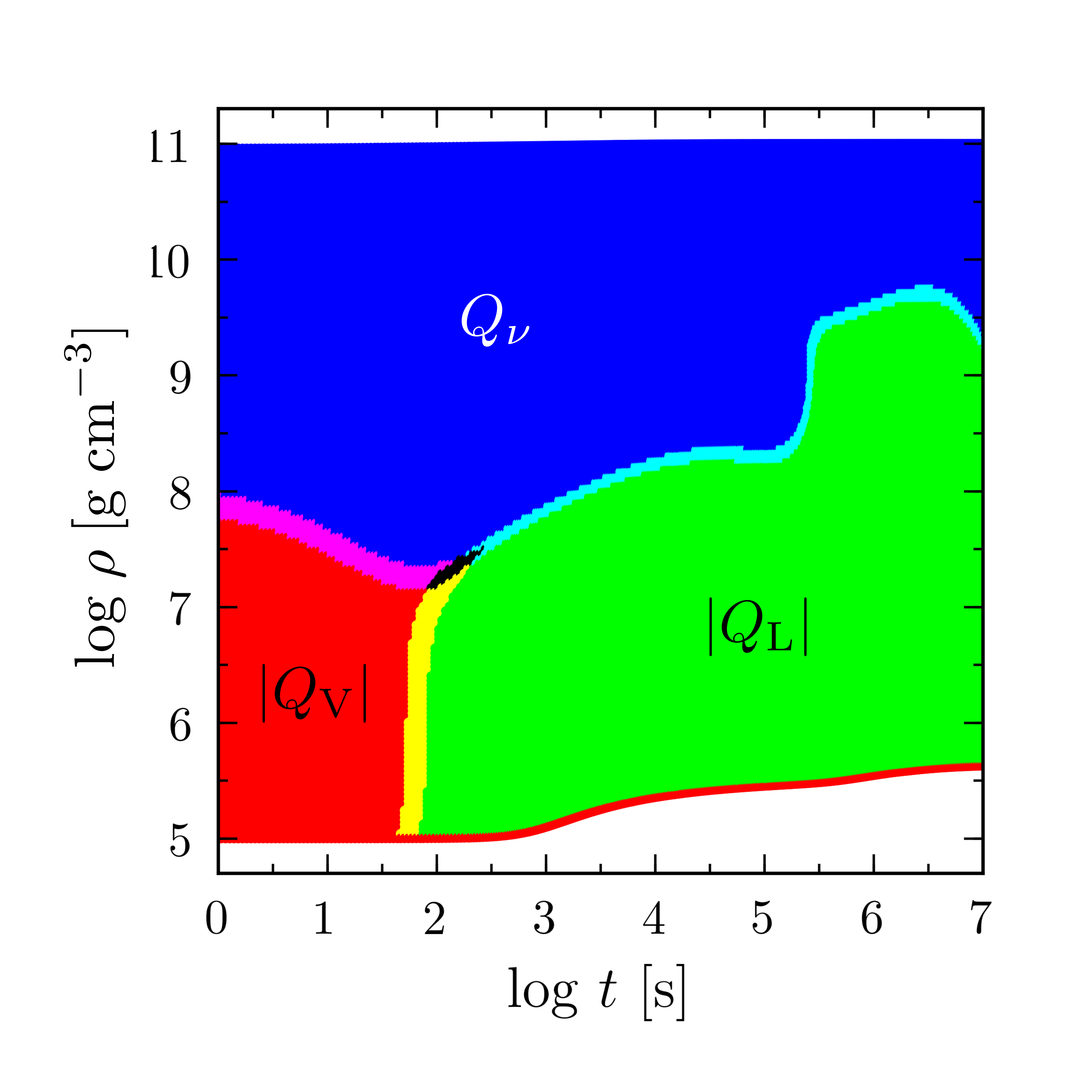}
\caption{Plot of dominant energy term in the energy balance \protect\Eq{Eq:Thermal_L} in model A:               
              blue for $Q_\nu$, red for $|Q_\mathrm{V}|$ and green for $|Q_\mathrm{L}|$; other colors are where two contribution are within 20\%
              of each other while in the black region all three are within 20\% of each other.
              (We use absolute values for quantities $Q_\mathrm{V}$ and $Q_\mathrm{L}$ that can be either positive or negative.)
              }
	\label{fig:Eprof1}
\end{figure}

\begin{figure}
	\includegraphics[height=8.0cm,keepaspectratio=true,clip=true,trim=0.0cm 0.5cm 0.7cm 0.6cm]{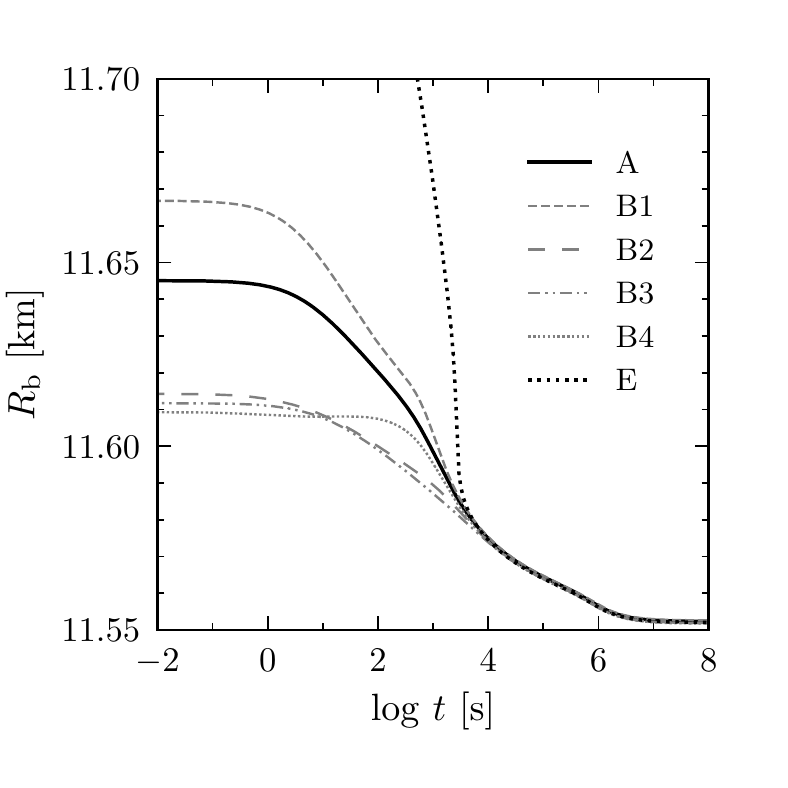}
	\caption{Time evolution of the radius of our $1.4 \, \Msun$ models. See text for details}
	\label{fig:Rad}
\end{figure}

\begin{figure}
	\includegraphics[height=8.0cm,keepaspectratio=true,clip=true,trim=0.0cm 0.5cm 0.7cm 0.6cm]{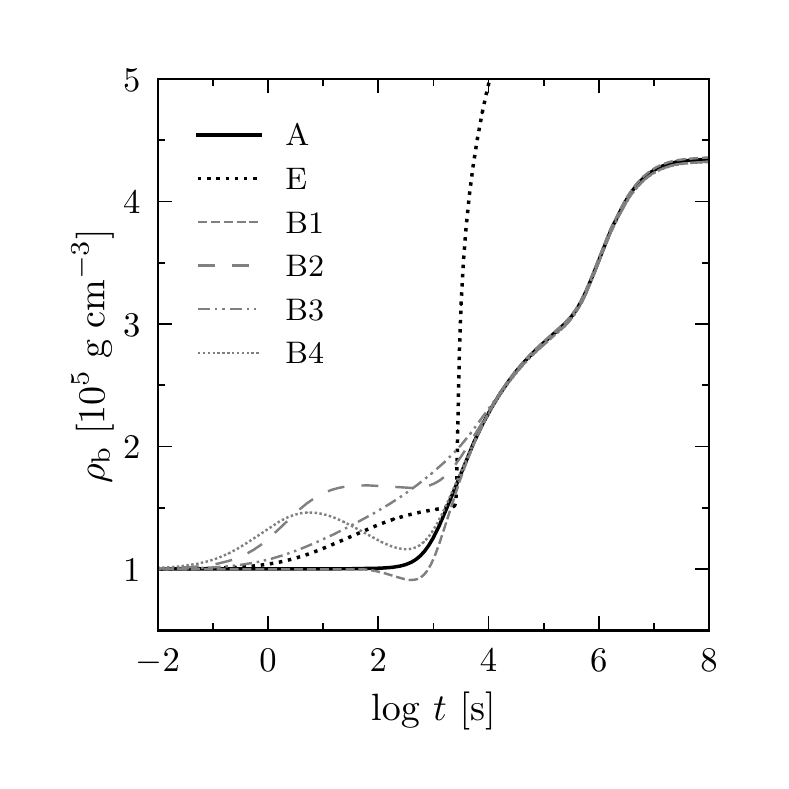}
	\caption{Time evolution of the boundary density $\rhob$ of our $1.4 \, \Msun$ models. See text for details}
	\label{fig:Rho}
\end{figure}

In the right panel of \Fig{fig:Cool1} we show a close up of the early evolution of $T_\mathrm{eff}^\infty$.
It is interesting to notice here that these cooling curves map their initial temperature profiles that were displayed in 
\Fig{fig:Initial}b: it results from a mapping of $T_0(\rho)$ into $T_\mathrm{eff}(t)$.
This mapping is good up to time $\sim 10^3$~s with $\rho$ up to $10^{8} \, \gcc$:
it was shown by \citet{Brown:2009aa} that as long as $T_\mathrm{eff}^\infty$ is controlled by heat transport from deeper layers
its value is determined mostly by the initial $T_0(\rho)$ at a depth whose thermal diffusion time scale to the surface is equal to the 
time elapsed from when this initial $T_0(\rho)$ was set.
In our case, the mapping ends when $t-\rho$ reaches a density where the evolution is driven by neutrinos more than by heat
diffusion toward the surface and this happens when approaching the phase ``2'' dominated by pair-annihilation neutrinos.
In \Fig{fig:Eprof1} we show the time evolution of the dominant energy term in the energy balance \Eq{Eq:Thermal_L} as a function of density
for model A: the details of such a plot are dependent on the assumed initial $T$ profile but that $Q_\nu$ eventually dominates at high densities (which,
as one can see, turn out to be above $\sim 10^8 \, \gcc$) is a simple result of the high $T$ dependence of neutrino processes and the strongly
raising $T$ profile as $\rho$ increases.

In the \Fig{fig:Rad} we show the evolution of the boundary radius $R_\mathrm{b}$ of our $1.4 \, \Msun$ models.
The different radii at early times are a direct reflection of the inner envelope temperature profiles: hotter envelopes are naturally more expanded.
Excluding the model E we find contractions of $R_\mathrm{b}$ of the order of 50 to 100 meters.
Similarly, in \Fig{fig:Rho}, we show the evolution of the outer boundary density $\rho_\mathrm{b}$ of the same models.
Since $\rho_\mathrm{b}$ evolves with $T_\mathrm{b}$ in such a way that 
$P_\mathrm{b}$ remains almost constant\,\footnote{
There is a small time evolution of $P$ in the outer layers because of contraction and the resulting small change in $g$, as seen, e.g., from \Eq{Eq:column}.},
and since $T_\mathrm{b}$ is
instantaneously correlated with $\Ts$ through the outer envelope, one sees that $\rho_\mathrm{b}$ is directly anti-correlated with $\Ts$
shown in \Fig{fig:Cool1}b.
On the contrary, $R_\mathrm{b}$ results form the integral of the thickness of underlying layers and its evolution is not directly correlated with 
the detailed evolution of $\rho_\mathrm{b}$ or $\Ts$ during phase ``1''.

\begin{figure}
\includegraphics[width=1.00\columnwidth]{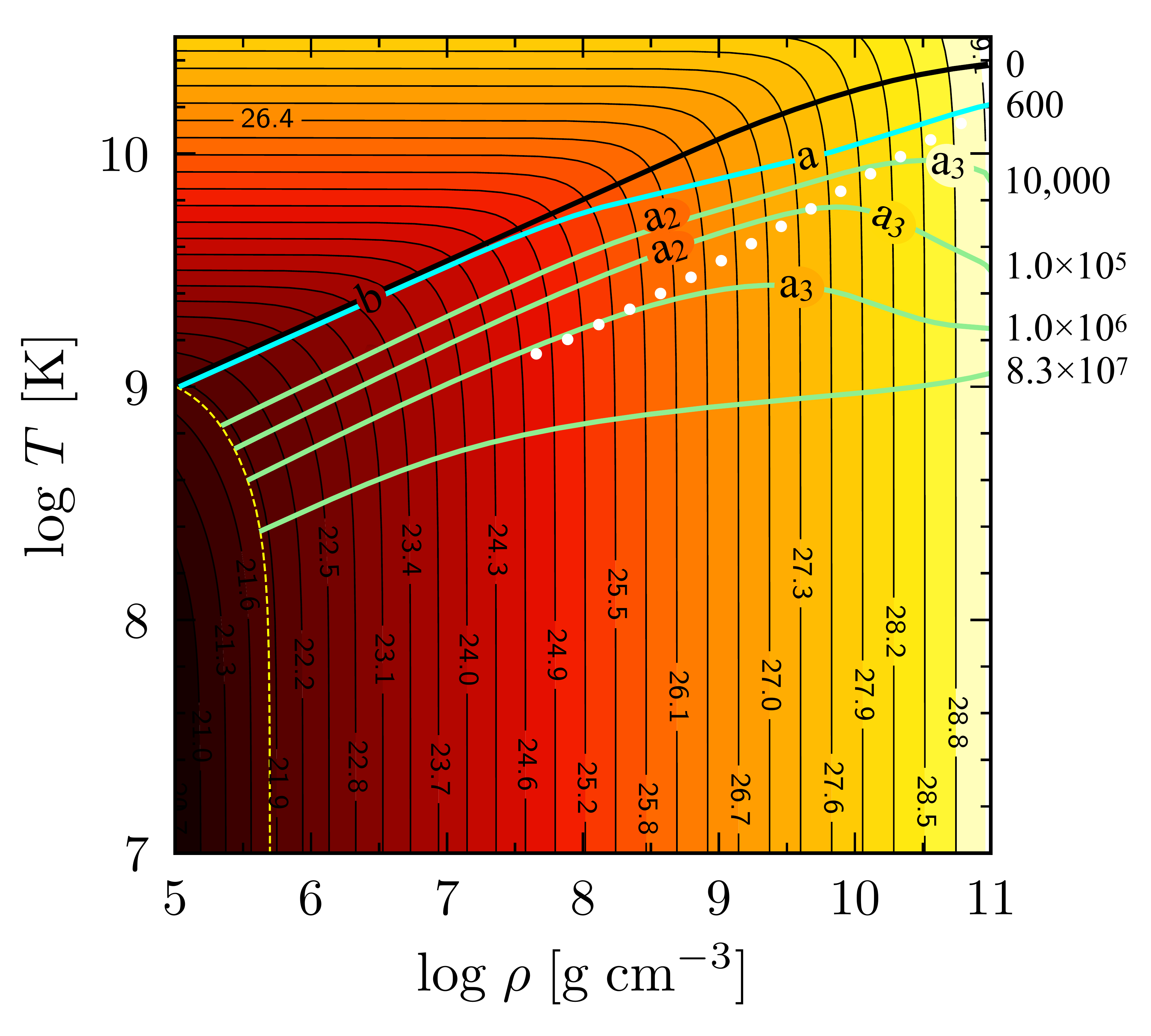}
\caption{Selected local temperature profiles of model~A.
	     Ages, in seconds, are indicated on the right margin.
	     Background color shows the pressure and contours are isobars labelled with decimal logarithm of pressure [in dyn~cm$^{-2}$].
	     The dashed (yellow) contour corresponds to the initial $P_\mathrm{b}$ and 
	     the thick dotted (white) line reproduces the one from \protect\Fig{fig:Nu_2D}.
	}
	\label{fig:Tprof_A}
\end{figure}

Considering, again, our model A in more detail as an illustrative case, we present in \Fig{fig:Tprof_A} a series of envelope temperature profiles.
Since we have no mass-loss in our models, the column density $y_\mathrm{a}$ of any layer is constant during the evolutions and hence each layer evolves
at (almost) constant pressure\,$^3$.
We display in the background of \Fig{fig:Tprof_A} the pressure of the medium and a series of isobars: matter evolves along these isobars during the cooling.
The profile at 600~s corresponds to the end of the early plateau during which $T_\mathrm{eff}$ is locked to $T_\mathrm{eff, Edd}$:
we can divide the inner envelope in two regions, layer ``a'' at densities above $\sim 10^8 \, \gcc$ where neutrino losses have had a significant effect
(compare with \Fig{fig:Eprof1}) and layer ``b'' below $\sim 10^8 \, \gcc$ where the temperature profile has almost not evolved.
During this phase $R_\mathrm{b}$ has decreased by some 40 meters (\Fig{fig:Rad}) but this is due to the contraction of layer ``a'' while 
layer ``b'' has not contracted but has rather been slowly sinking, keeping its initial density and temperature profile.
After this first phase $T_\mathrm{b}$ and the whole layer ``b'' are thermally connected to the temperature in layer ``a'':
$T_\mathrm{b}$, and $\Ts$, begin to drop following the cooling of ``a''.
As a result the layer ``b'' begins to contract and $\rho_\mathrm{b}$ to increase as exhibited in \Fig{fig:Rho}.
The evolution of the temperature profile from 600 up to $10^6$~s shows a clear difference between the region dominated by pair neutrinos, 
layer a$_2$, versus plasmon neutrinos, layer a$_3$, (separated in the figure by the (white) dotted line).
As time runs the layer a$_2$ encompasses less and less, while layer a$_3$ encompass more and more, mass.
(At these phases layer ``b'', whose energetics is dominated by either $Q_\mathrm{V}$ or $Q_\mathrm{L}$, always start at densities around 
$10^8$ as seen in \Fig{fig:Eprof1}.)
This different evolution of layers a$_2$ versus a$_3$ is easily understood by considering the difference in temperature dependence of these two neutrino processes (see \Fig{fig:Nu_2D})
that result in the strongly different cooling time scales displayed in \Fig{fig:Scales}a.
As long as part of the envelope is in the pair neutrino regime this layer a$_2$ will drive the evolution of the outer layers and we are in phase ``2''
while after $\sim 10^6$ the layer a$_2$ has disappeared , the cooling of the outer layers is driven by layer a$_3$ and we entered phase ``3''.
It is interesting to see in \Fig{fig:Eprof1} that at age $\sim 10^{5.5}$~s,  when the cooling curve passes through the ``knee'', the decrease in the pair neutrino
emission is so strong that the energetics of layers that were previously dominated by $Q_\nu$ are now dominated by $Q_\mathrm{L}$
up to densities of $10^{9.5} \, \gcc$.

\begin{figure}
\includegraphics[width=1.00\columnwidth]{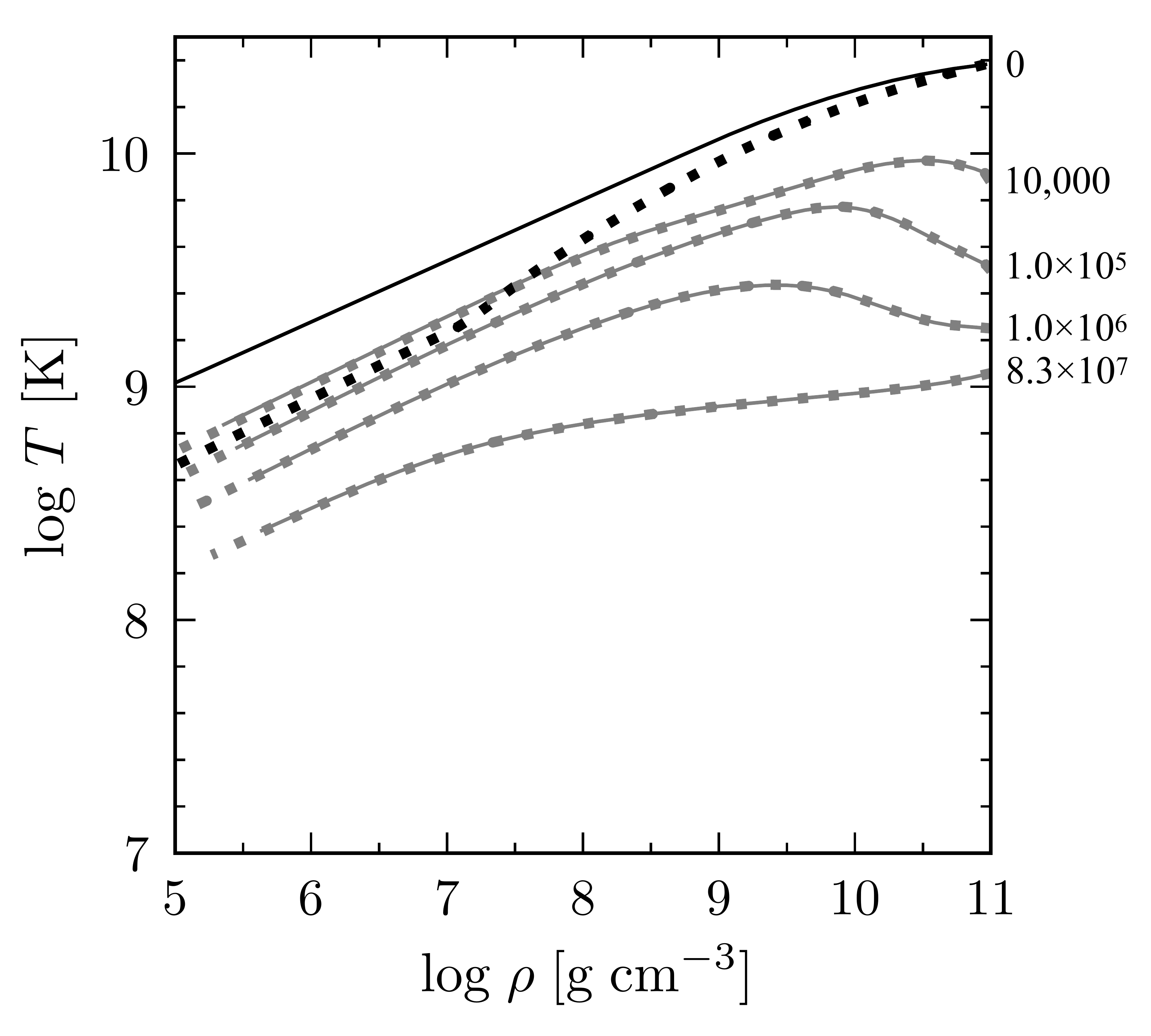}
\caption{Selected local temperature profiles of  model~F (dotted lines) compared to model A (solid lines). Ages, in seconds, are indicated on the right margin.}
	\label{fig:Tprof_AF}
\end{figure}

\subsection{Robustness of our $1.4 \, \Msun$ Star Results.}
\label{sec:robust}

After this thorough study of our model A let us have a look at models B1 to B4.
They are all based on the same two starting points, an interior initially at a temperature $T_0 \simeq 2.5 \times 10^{10}$ K as implied by studies of
proto-neutron stars and binary mergers, and a surface luminosity initially at $L_\mathrm{Edd}$, but they have different $L$ and $T$ profiles 
in-between.
These four scenarios have different evolutions only during the early relaxation phase ``1'' as clearly seen in Figs.\ \ref{fig:Cool1}, \ref{fig:Rad}, and
\ref{fig:Rho}, a phase where $\Ts$ is driven by the heat diffusion in the low density part of the inner envelope.
However, once the cooling is controlled by neutrino emission, phases ``2'' and later ones, their evolutions are identical to scenario A:
neutrinos are so efficient that they rapidly erase any remembrance of the initial conditions.
Nevertheless, during the first $10^3$ s these different scenarios only span a range of $\Ts^\infty$ between $1.6$ to $2.1 \times 10^7$ K 
and a surface luminosity $L^\infty$ between $1$ to $3 \times 10^{38}$ erg s$^{-1}$.
Hence, {\em we have very similar luminosity evolutions during the first half an hour and then a basically universal evolution for the first year}.
Notice that the neutrino processes from either pair annihilation and plasmon decays depend only on the temperature and the electron density
and do not depend on the type of nuclei present in the medium. 
It is only later, during phase ``4'' controlled by neutrino emission from electron-ion bremsstrahlung that the actual chemical composition of
the medium becomes important.

On the other side, it is well known that the chemical composition of the outer envelope has a strong effect on the 
$T_\mathrm{b} - T_\mathrm{s}$ relationship, lighter elements having a larger thermal conductivity and resulting in higher $T_\mathrm{s}$
for a given $T_\mathrm{b}$. 
How large is this effect and how likely is the presence of light elements in the high temperature envelope we employ is an open question.
Notice that at densities $\sim 10^5 \, \gcc$ and temperatures $\sim 10^9$ K thermonuclear rates are enormous and the survival of light elements is doubtful.
We intent to tackle these issues in a forthcoming work.

As a distinct initial configuration let us consider our model F which started with the same 
$T_{\mathrm{c},0} = 2.5 \times 10^{10}$~K at density $\rhoc = 10^{11}~\gcc$ but a much lower outer boundary temperature 
$T_{\mathrm{b}, 0} \simeq 0.4\times 10^9$~K at $\rhob = 10^{5}~\gcc$.
This model started with a clearly sub-Eddington $L^\infty$ as seen in \Fig{fig:Cool1} but after a few hundreds seconds its surface
layers heat up because of the high flux coming from the inner envelope.
In \Fig{fig:Tprof_AF} we show the temperature profile evolution in the inner envelope:
compared to model A the initial profile has no choice but to have a stronger gradient in the inner part in order to reach a lower $T_{\mathrm{b}, 0}$
and this is the cause of the rise in $\Ts$ at later times when this enhanced flux reaches the surface.
As in the other models, after $\sim 10^4$ seconds the temperature profiles have converged to the universal profiles and are indistinguishable from
the ones of model~A.

As described in \App{app:NCv} there is some uncertainty regarding the nuclei contribution to the specific heat,
but it is only relevant at densities above $10^{10}~\gcc$ and temperatures well above $10^9$ K.
This implies that this uncertainty has no effect on the duration of the initial Eddington luminosity phase:
this phase terminates when neutrino cooling in region a$_2$ (see \Sec{sec:1.4Msun}) drives the evolution of $\Ts$
and in this region the nuclei specific heat is negligible.

\subsection{Evolution of High and Low Gravity Stars}
\label{sec:others}

\begin{figure}
	\includegraphics[height=8.0cm,keepaspectratio=true,clip=true,trim=0.25cm 0.5cm 0.5cm 0.6cm]{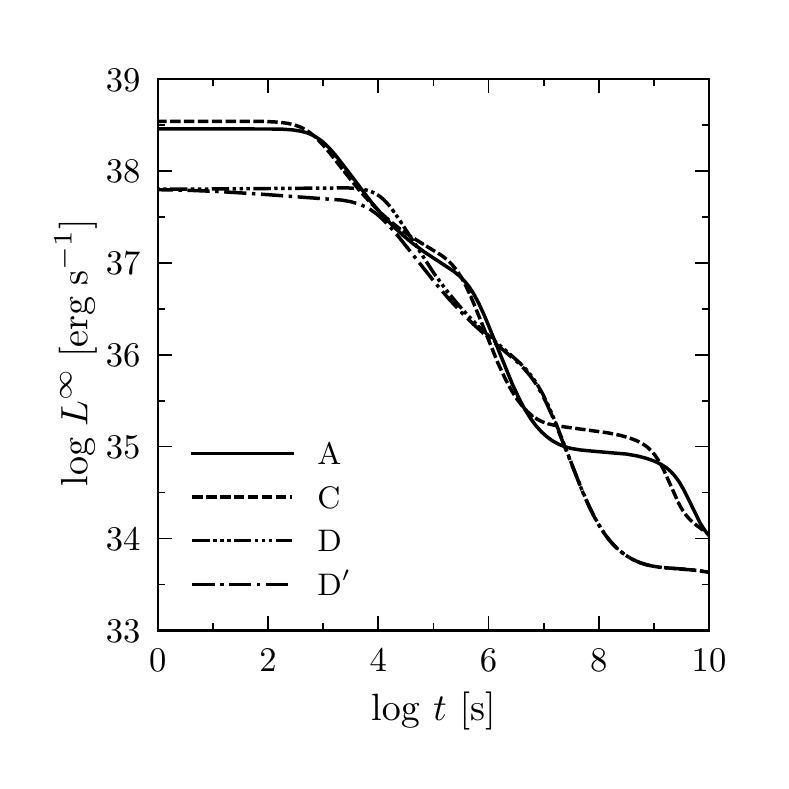}
	\caption{Cooling curves of our models A ($1.4 \, \Msun$), C ($2 \, \Msun$), D, and D$^\prime$ ($0.25 \, \Msun$).
	See details in the text.}
\label{fig:Cool2}
\end{figure}

In the \Fig{fig:Cool2} we show the cooling curves for different gravities. 
Model C with $g_\mathrm{s, 14} \simeq 3.2$ turns out to be very similar to model A with $g_\mathrm{s, 14} \simeq 1.6$:
we are plotting the red-shifted luminosity and its intrinsically higher luminosity is in large part compensated by a higher red-shift.
For the low gravity models, D and D$^\prime$, the lower Eddington luminosity clearly shows and moreover the relaxation time is much longer:
the initial relaxation phase ``1'' last much longer and an Eddington luminosity can be sustained for more than $10^4$~s
versus less than $10^3$~s for models A and C.
As a curiosity, in model D$^\prime$ we have arbitrarily switched-off the contraction energy of \Eq{Eq:QV}:
as a result during the initial relaxation phase the luminosity slightly decreases instead of staying almost constant as in model D.
Nevertheless, at ages between $10^4$ up to $10^7$ s (i.e., between three hours up to three months), 
luminosities of these three models with very different surface gravities
are still very similar and only actually differ in details (as, e.g., the time at which the ``knee'' occurs).

\subsection{At the crossroads of different physical regimes}
\label{sec:crossroad}

We finally describe our model E which has a surface luminosity twice higher that $L_\mathrm{Edd}$ implying a significant mass-loss.
However, in the inner envelope the luminosity in this model is still sub-Eddington due to the fact that the opacity $K$
is much lower in this region than at the photosphere.
We can thus still model the inner envelope within our quasi-static formalism.
As seen in \Fig{fig:Cool1} this super-Eddington phase can last longer than the Eddington phase of our other models:
about 2,600~s after which time $\Ts$ suddenly drops.

\begin{figure}
\smallskip
\includegraphics[width=1.0\columnwidth]{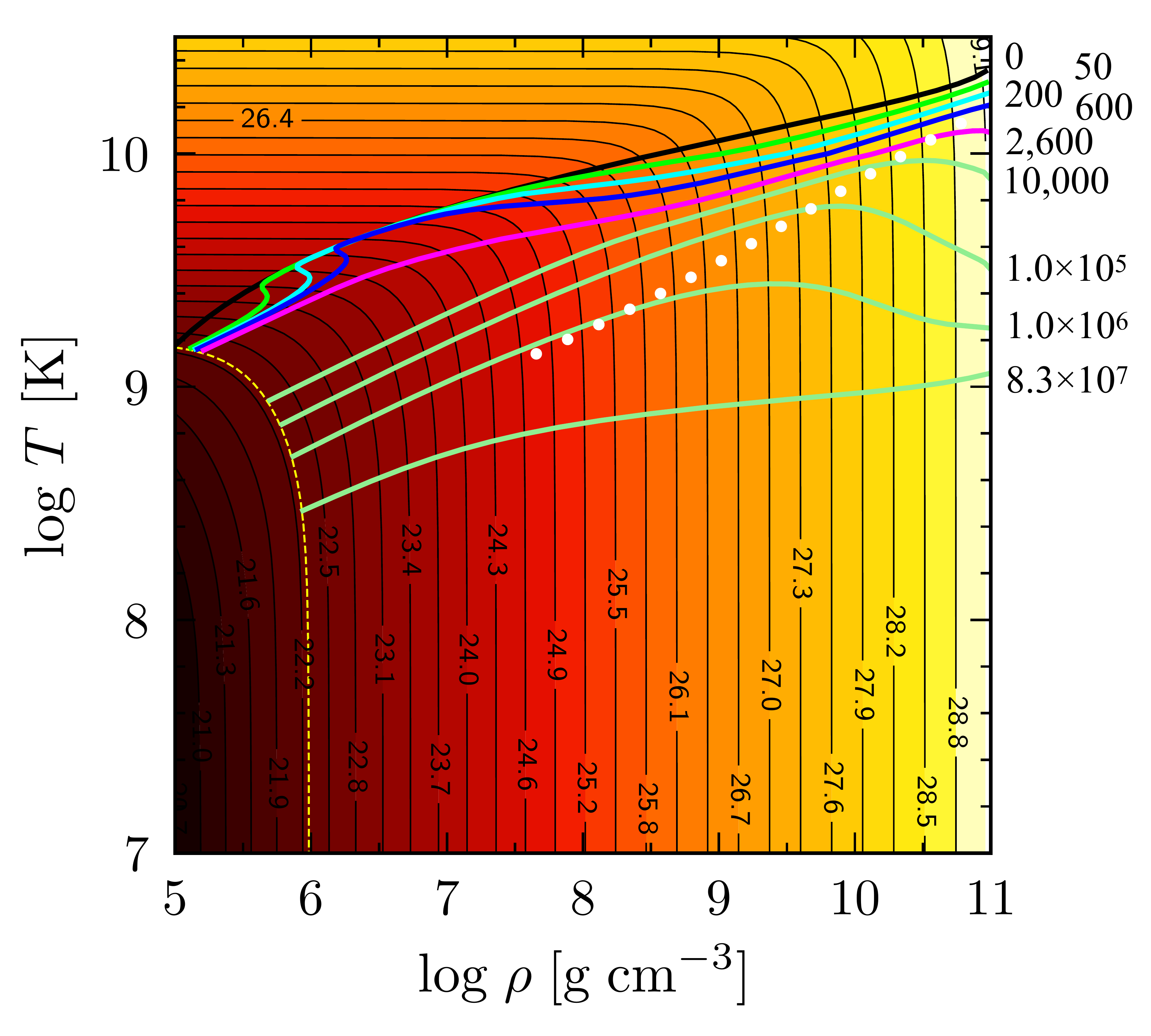}
\caption{Selected local temperature profiles of model~E.
	     Ages, in seconds, are indicated on the right margin.
	     Background color shows the pressure and contours are isobars labeled with $\mathrm{Log} \, P$ [dyn~cm$^{-2}$].
	     The dashed (yellow) contour corresponds to the initial $P_\mathrm{b}$.
	}
	\label{fig:Tprof_E}
\end{figure}

In \Fig{fig:Tprof_E} we illustrate the evolution of this model through it $T$-profiles.
Notice that at early times the low-density part of the inner envelope is clearly in the radiation/pair dominated regime.
This regime corresponds in this figure to the region where the isobars are horizontal, i.e., $\rho$ independent,
with $P \propto T^4$.
In contradistinction, in all our other Eddington luminosity models the inner envelopes were always in a regime where matter
made a strong contribution fo the pressure as can be seen, e.g., in \Fig{fig:Tprof_A}.
This super-hot model E results in a strongly puffed-up envelope, because of radiation pressure, as seen from the larger radius $R_\mathrm{b}$ in \Fig{fig:Rad}.
The first four $T$-profiles displayed in \Fig{fig:Tprof_E} show a rapid contraction at the lowest densities.
This contraction occurs at (nearly) constant pressure\,$^3$, hence at (nearly) constant temperature maintaining a (nearly) constant $\Ts$, and the gravitational energy 
released by this contraction is used to power the super-Eddington surface luminosity. This contraction wave propagates inward until it reaches the cooling wave from the neutrino cooling propagating outward from the denser regions.
After $\sim 2,600$~s further evolution along the isobars implies a significant temperature drop, the inner envelope
entering a different pressure regime, and the end of the super-Eddington phase.
After $\sim 10^4$~s this model has forgotten its initial configuration and follows the same evolution as all our other $1.4\, \Msun$ Eddington luminosity models.

\subsection{Effects of a Strong Magnetic Field}
\label{sec:magnetic_field}

Our neo-neutron star study assumes spherical symmetry and neglect the effect of both fast rotation and the presence of a strong magnetic field
and is, thus, only a very first step in this direction.
Since most realistic scenarios where a neo-neutron star may be observable are likely to produce a fast rotating and strongly magnetized star
and it is imperative to estimate expectable deviations of realistic models from our idealized ones.
A simple model intended to mimick fast rotation with a low gravity model was presented in subsection \ref{sec:others}.
The case of a strong magnetic field is more involved as it not only breaks spherical symmetry, but introduces strong anisotropies in
almost every physical ingredient of our models at both the micro- and macroscopic levels, and modifies the equation of state, opacities, thermal conductivities and also possibly neutrino emission
(for reviews about magnetic field effects we refer the reader to 
\citealt{Yakovlev:1994aa} and \citealt{PPP15}).

The heat transport anisotropy in the presence of a strong magnetic field and its effects on the thermal emission have been amply studied 
for a long time in neutron star envelopes (starting from, e.g., \citealt{Greenstein:1983xu} and \citealt{Page:1995yb}) and also in deeper layers
of the crust (see, e.g., \citealt{Geppert:2004ep,Geppert:2006zt} and \citealt{Perez-Azorin:2006rq}).
The overall effect is that regions of the surface where the magnetic field is tangential to it will be much colder that regions where the field is radial.
Due to quantum effects on the thermal conductivity the hot regions with a radial field are moreover slightly hotter than they would be in the absence of
a magnetic field.
As a result, when integrating the outcoming flux over the whole stellar surface, one generally obtains luminosities similar to the 
non magnetic case with uniform surface temperature (see, e.g., \citealt{Page:1996aa} for examples with dipolar+quadrupolar magnetic field
geometries).
We, thus, do no expect that the magnetically induced anisotropy in heat transport results in significant deviations from our results.

A second point of interest is the duration of the early Eddington luminosity 
phase.
It is controlled by neutrino emission from the pair annihilation process which occurs
at densities above $\sim 10^8 ~ \gcc$, the layer ``a'' in \Fig{fig:Tprof_A} and 
clearly illustrated by the difference between the models
A and A$^\prime$ in panel (a) of \Fig{fig:Cool1}.
This process is not affected by magnetic field at least up to a strength $\sim 
10^{14}$~G \citep{Kaminker:1994js}.
It is only for much stronger fields, which become strongly quantizing even at 
temperature above $10^{10}$~K in this density range,
that the pair annihilation process may increase the neutrino losses and reduce the duration of an Eddington phase,
but unfortunately there is no complete calculation available in this density-temperature range to confirm this statement.
Synchrotron neutrino emission cannot compete at very high temperatures with 
the pair annihilation process, at least for fields below $10^{15}$~G, 
and thus is not expected to affect the Eddington luminosity phase for these magnetic field intensities.
However, in later phases dominated in the non magnetized cases by the plasmon decay process (phase ``3'' shown in panel (a) of \Fig{fig:Cool1})
synchrotron neutrinos will increase losses when the field is above $\sim 
10^{14}$~G and accelerate the cooling.

The third interesting effect of the presence of a strong magnetic field is the strong suppression of the opacity for extraordinary mode (XO) photons,
$\kappa_\mathrm{XO}(\omega) \approx (\omega/\omega_c)^2 \kappa_\mathrm{Th}$ when $\omega \ll \omega_c$,
where $\omega_c$ is the electron cyclotron frequency and $\kappa_\mathrm{Th}$ the electron-scattering (Thomson) opacity.
As a results, when mode switching between the XO and the ordinary (O) mode is taken into account, the critical luminosity $L_c$ (i.e., the Eddington
luminosity in the presence of the magnetic field) is increased compared to the zero field $L_\mathrm{Edd}$ of \Eq{Eq:LEdd} by
$L_c \approx (\omega_c/\omega) L_\mathrm{Edd}$ and can easily reach 
$10^{41 - 42}$~erg~s$^{-1}$ \citep{Miller:1995aa}.
Even higher luminosities, $\sim 10^{44}$~erg~s$^{-1}$, have been observed 
during the magnetar Giant Flares (see, e.g., \citealt{Kaspi:2017us})
where a hot plasma is likely confined in the magnetosphere \citep{Thompson:1995cd}.
How long could a strongly magnetized neo-neutron star sustain a high thermal 
luminosity close to $L_c \sim 10^{41 - 42}$~erg~s$^{-1}$ is the crucial 
question.
In terms of energetics, our model A sustained $L\simeq 3 \times 
10^{38}$~erg~s$^{-1}$ for about 1000 second and emitted a total of $\sim 3 
\times 10^{41}$~ergs
while our extreme model E kept $L\simeq 6 \times 10^{38}$~erg~s$^{-1}$ for 
about 2500 second emitting a total of $\sim 2 \times 10^{42}$~ergs:
this second model had a much hotter and bloated envelope that contained much more energy, both gravitational and thermal,
allowing it to sustain a higher luminosity for a long time.
Whether a strongly magnetized neo-neutron star envelope will be hot enough and contains sufficient energy to sustain a luminosity close to its
possible $L_c$ for a long time (a significant fraction of an hour) is difficult to assess at present time and definitely requires a detailed calculation with
all the appropriate physics included. 
We hope to address this issue in a future paper.

\section{Observational Prospects}
\label{sec:obs}

\subsection{Core Collapse Supernova and Supernova Remnants}

\begin{figure}
\smallskip
\centerline{\includegraphics[width=1.0\columnwidth]{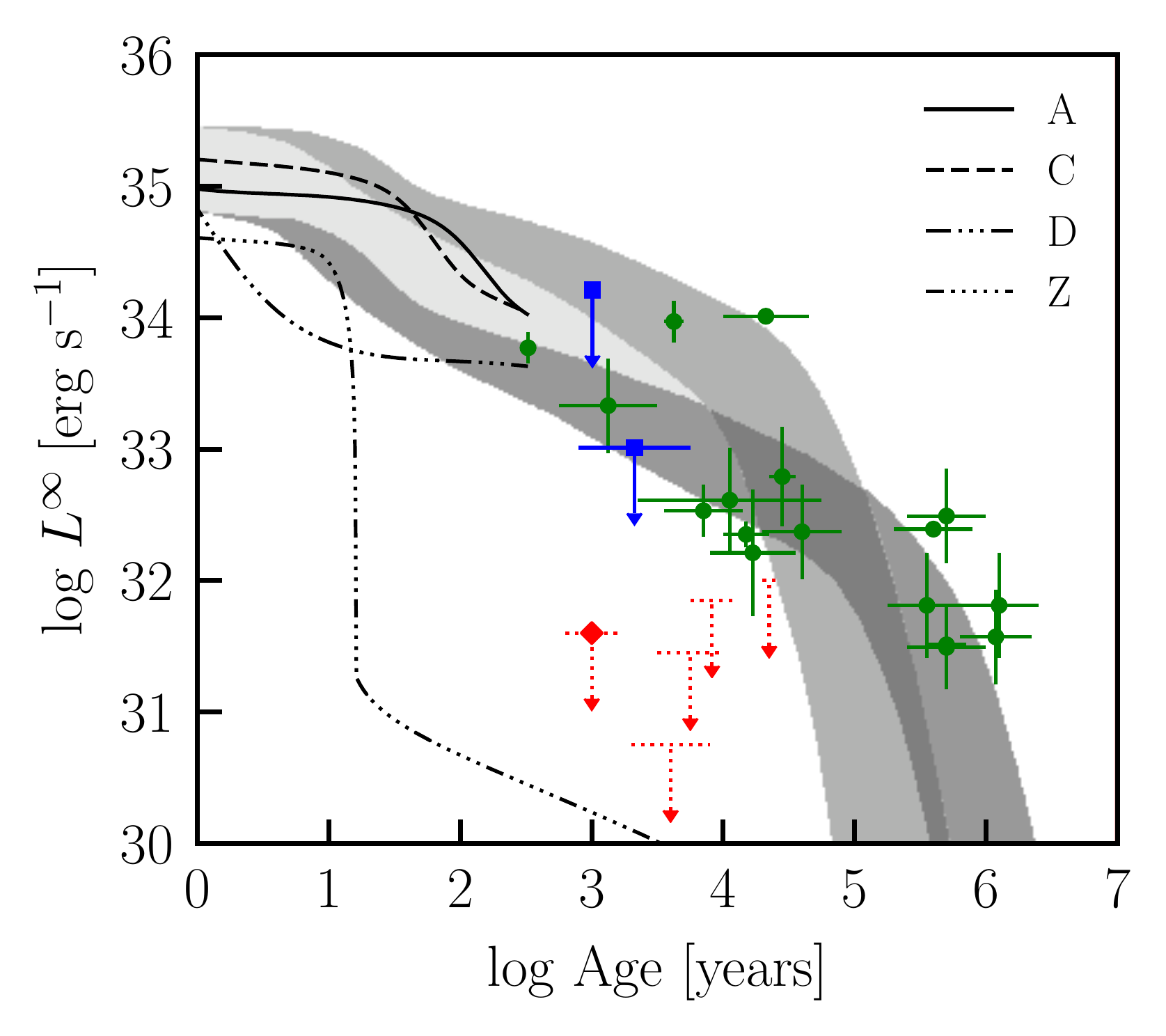}}
\caption{Plot of luminosity as a function of the age of the remnant for nearby neutron stars together with the predictions from  neutron star cooling models.
The (green) dots are measurements and the (blue) squares are upper limits from detected neutron stars (data taken from \citealt{BY15})
while the (red) dotted error bars are upper limits on the compact objects, black holes or neutron stars, expected to be presents in five core collapse supernova remnants. 
These remnants are, in order of increasing age: G043.3--0.2 (a.k.a. W49B, marked by a diamond) from \citet{2013ApJ...764...50L}, 
G127.1+0.5, G084.2+0.8, G074.0--8.5, and G065.3+5.7 from \citet{Kaplan:2004uv,Kaplan:2006fs}.
Shaded areas show model predictions of \citet{PLPS04,Page_etal09} for the  minimal cooling of neutron stars that cover uncertainties on the chemical composition of the
envelope and nucleon pairing at high densities.
In contradistinction the dashed-pentadotted curve, Z, exemplifies the effect of fast neutrino emission from the direct Urca process expected to occur in massive neutron stars \citep{Boguta:1981jz,LPPH91} resulting in very cold stars \citep{Page:1992cp}.
Also plotted are the three different models (A, C, D) shown in \Fig{fig:Cool2}.}
	\label{fig:snr}
\end{figure}

Neutron stars are great laboratories for studying  the equation of state of nuclear-density matter. The study of supernova remnants, on the other hand, help us elucidate  the composition and structure of their stellar progenitors \citep[e.g.][]{2011ApJ...732..114L}. By associating neutron stars with  supernova remnants, we can obtain unique information  about
these systems  that is unavailable when we study them  separately.  What is more,  supernova associations provide a way to 
independently constrain the age of the neutron star as well as  searching for 
former binary surviving companions.

There are, however, clear limitations  
that  prevent us from uncovering young systems;  the most noticeable being that the ejecta  gas needs to be transparent to the neo-neutron star radiation.
For optical (in the absence of dust) and  high X-ray energies ($\gtrsim 10$ keV), electron scattering provides the main opacity \citep{1970ApJ...162..737B}. 
Let's consider a cloud of gas with mass $M_{\rm ej}$ ejected from the explosion.
The cloud radius expands freely as $R = v_{\rm ej}t$ where
$v_{\rm ej} = \sqrt{ 2E_{\rm ej}/M_{\rm ej}}$ is the characteristic velocity,  $t$ is the time since
ejection and $E_{\rm ej}$ is  the total kinetic energy.  We thus expect an homogeneous envelope to become transparent after a time   
\begin{equation}
t_{\tau=1} \approx  0.5 \left({E_{\rm ej} \over  10^{51}{\rm erg}}\right)^{-1/2} \left({M_{\rm ej}\over 5\, \mathrm{M}_\odot}\right) \;{\rm yrs},
\end{equation}
which is frustratingly about the duration of the Eddington luminosity phase.
Here we assume $\kappa \approx 0.1$~cm$^2$~g$^{-1}$, which is a reasonable value  for ordinary supernova material. At lower X-ray energies ($\epsilon_{\rm x}\lesssim$ 3 keV), photoionization may delay the time at which the envelope becomes transparent by an additional factor of $\sim 5 ({\rm 
\epsilon_{\rm x} /1 keV})^{-3/2}$.

 Despite this, the youngest neutron star we have  uncovered has an age of about 340 years \citep{Fesen:2006aa}.
In a few instances, a surviving binary companion has been detected in post-explosion deep optical  imaging of extragalactic supernova \citep{2009Sci...324..486M,2014ApJ...793L..22F}. In the case of 1993J, for example,  the brightness of the  transient dimmed sufficiently after about a  decade so that its spectrum showed the features of a massive star superimposed on the supernova \citep{2004Natur.427..129M}. 

It is perhaps a stinging fact that despite the expected  manifestations of  neo-neutron stars, one of the main issues in the field has been
that most Galactic supernova remnants have no detectable central source.   \Fig{fig:snr} shows the current detections and upper limits of thermal emission in nearby neutron stars with model predictions.
Observational selection effects are clearly at play when uncovering young objects yet there is the possibility  that a sizable 
 fraction of massive star collapses  might produce black holes rather than neutron stars, with the clearest example being W49B \citep{2013ApJ...764...50L}
 and the other four examples, from \citet{Kaplan:2004uv,Kaplan:2006fs} all plotted in \Fig{fig:snr}.

\subsection{Neutron Star Mergers, Short Gamma-Ray Bursts and LIGO Events}

The merger of binary neutron stars
and the subsequent production of a beamed, relativistic outflow is believed to trigger short gamma-ray bursts \citep[e.g.][]{2017ApJ...848L..13A}
and expel metallic, radioactive
debris referred to as a kilonova \citep[e.g.][]{2017Natur.551...80K}.  The ultimate  fate of the post-merger remnant remains unclear and is dependent  on the
mass limit for support of a hot, differentially rotating neutron star.  The merged remnant can  either collapse and form  a low-mass black hole \citep{1989Natur.340..126E,2011ApJ...732L...6R,2014ApJ...788L...8M} or survive as a hyper-massive neutron star \citep{1992Natur.357..472U,Kluzniak:1998uq,2008MNRAS.385.1455M, 2018ApJ...856..101M}. In that case, the detectability of the hyper-massive remnant will depend primarily on  the orientation of the merging binary.

\begin{figure}
\smallskip
\includegraphics[width=1.0\columnwidth]{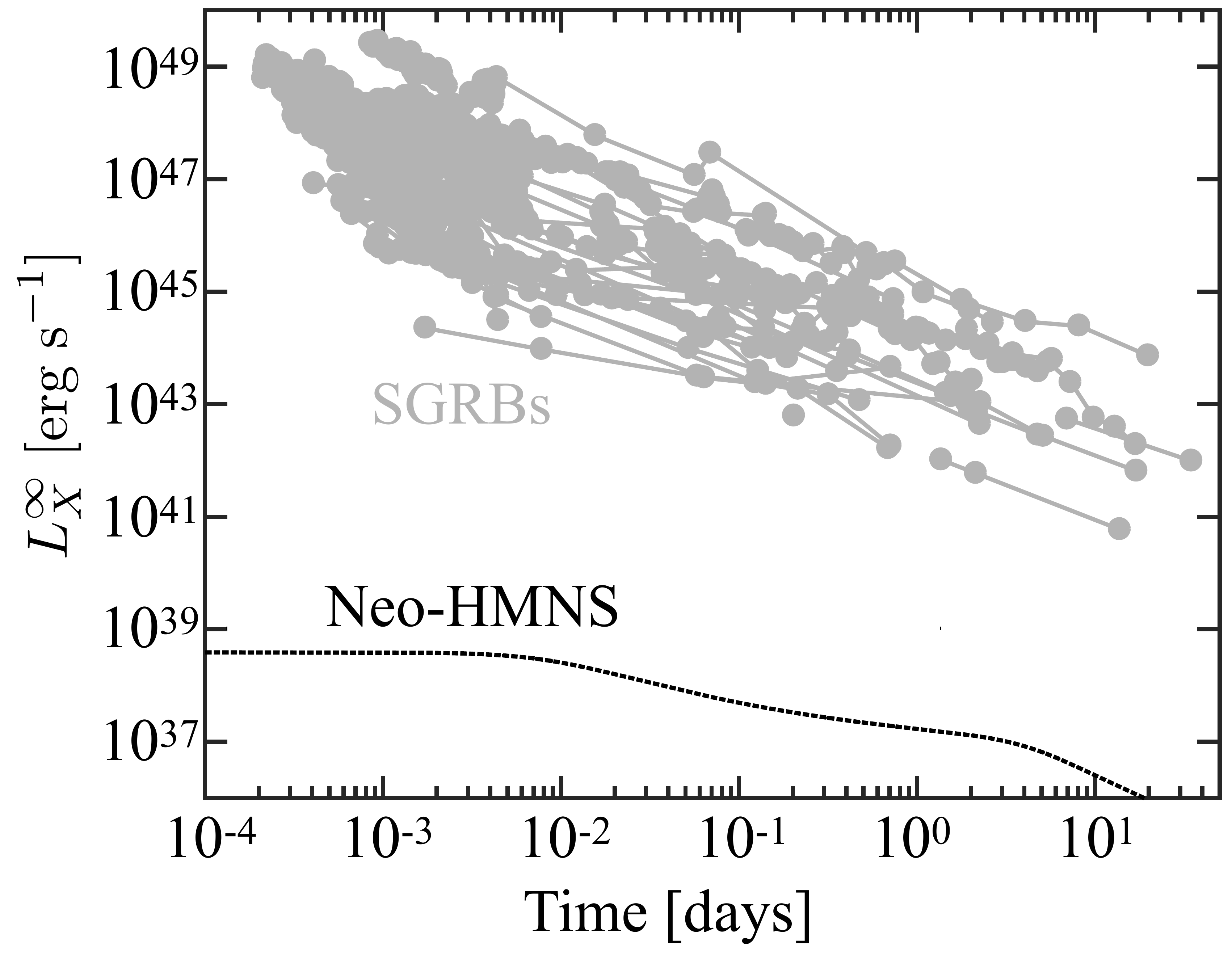}
\caption{Plot of the on-axis X-ray afterglow light curves from a sample of 36  short GRBs with well sampled  light curves and redshifts compiled by \citet{2017ApJ...848L..23F}. Also plotted is the X-ray luminosity of the cooling {\it model C} shown in \Fig{fig:Cool1},  labeled as hyper-massive neutron star (HMNS). }
	\label{fig:sgrbs}
\end{figure}

When the relativistic jet points in the direction of the observer, the event will likely be detected as a classical short gamma-ray burst \citep{2007PhR...442..166N,2007NJPh....9...17L} and the X-ray emission emanating from the surviving  remnant will  be buried by  the  luminous afterglow emission. This can be seen in \Fig{fig:sgrbs}, where we compare the X-ray  luminosity of the cooling {\it model C} shown in \Fig{fig:Cool2} to the on-axis X-ray afterglow luminosities of a sample of short gamma-ray bursts.

\begin{figure}
\smallskip
\includegraphics[width=1.0\columnwidth]{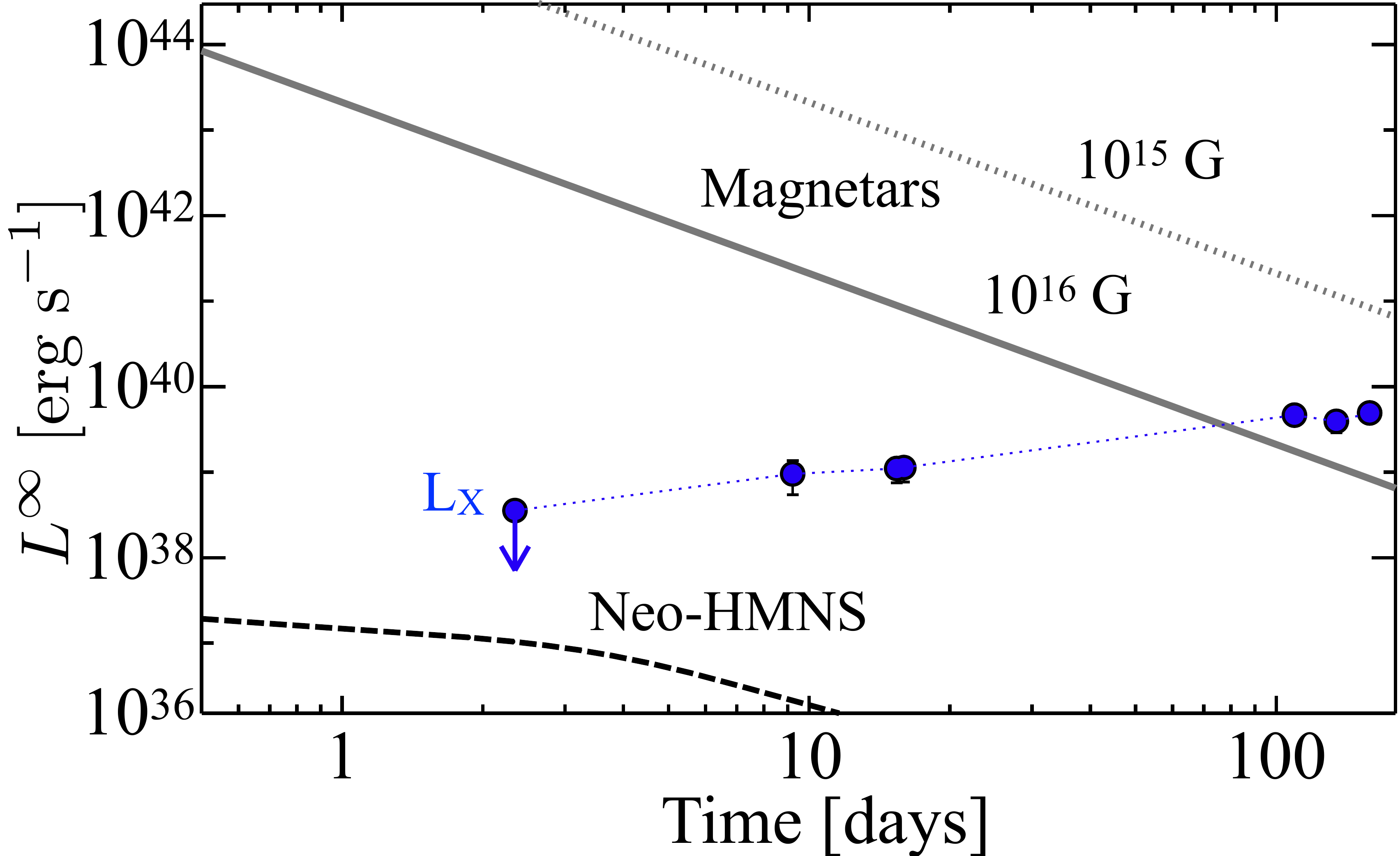}
\caption{Plot of the  X-ray
light curves  of the  counterpart to GW170817 from {\it Chandra} (0.3-10 keV) and  cooling {\it model C} shown in \Fig{fig:Cool1} (labeled as Neo-HMNS).  Also plotted, for comparison,  are the spin-down (bolometric) luminosities expected for a stable hyper-massive magnetar.  Adapted  from \citet{2018ApJ...856L..18M}. }
	\label{fig:gw170817}
\end{figure}

Our ability  to directly uncover the emission of the newly formed, hyper-massive neutron star drastically increases when the event is off-axis, 
as was the case for GW170817 \citep{2017ApJ...848L..13A,2017ApJ...848L..12A}. 
In August 2017,  \citet{2017Sci...358.1556C} discovered the first optical counterpart to a
gravitational wave source.  In this case, the cataclysmic merger of two 
neutron stars.
This landmark discovery initiated
the field of gravitational wave astronomy and enabled  an exhaustive observational campaign \citep{2017ApJ...848L..12A}. 
In \Fig{fig:gw170817} we show the luminosity of the X-ray counterpart to  
GW170817 \citep{2018ApJ...856L..18M} 
that is seen to be significantly dimmer than the one expected from  the 
spin-down of a highly-magnetized, rapidly rotating  remnant but  only slightly 
brighter than the X-ray luminosity predicted for  the relevant cooling {\it model 
C} that is plotted  in \Fig{fig:Cool2}. 

\begin{figure}
\smallskip
\includegraphics[width=1.0\columnwidth]{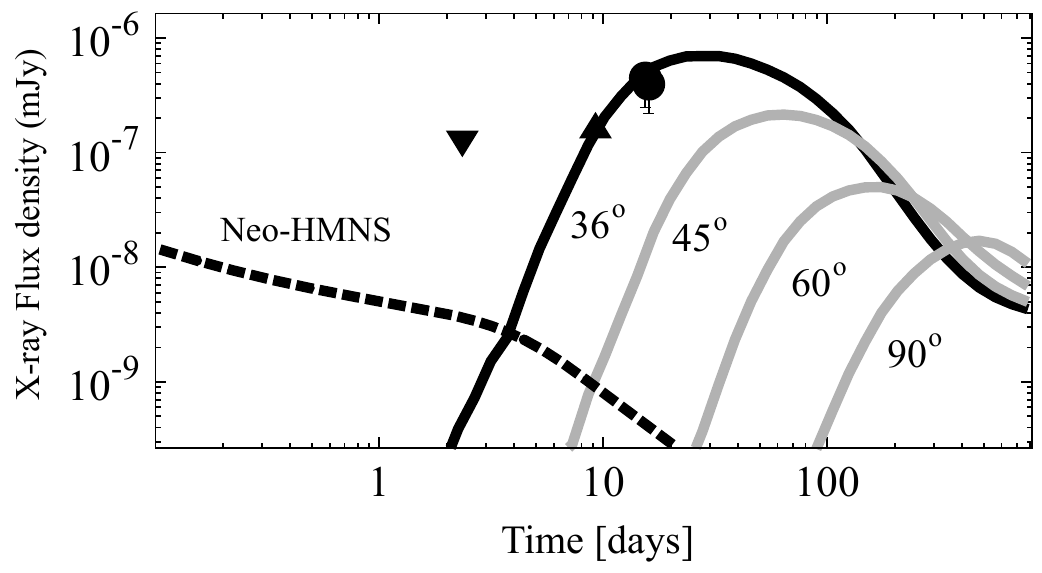}
\caption{Plot of the  X-ray
flux density at 1 keV  for an off axis model with $10^{49}$ erg and $\theta_{\rm obs}=36^{\rm o}$ from  \citet{2017ApJ...848L..20M} aimed at  providing a reasonable description of the X-ray data of  GW170817. 
Also shown are the corresponding  models 
for observers seeing the same event  but further away from the axis of the jet and the cooling {\it model C} (labeled as Neo-HMNS).}
	\label{fig:offaxis}
\end{figure}

The constraints imposed by the  afterglow observations favor the idea that 
GW170817  was a typical GRB jet seen off axis  
\citep{2017ApJ...848L..13A,2017ApJ...848L..12A}. This interpretation 
assumes that our line of sight is tens of degrees from the core of the jet and thus
suggests that the prospects for detecting the remnant directly might be doable
for future events, in particular if they are seen further away from the  axis of
the jet as can be seen in \Fig{fig:offaxis}.

As discussed in the case of core collapse supernova, one of the challenges  for direct detection is that the neo-neutron star  is  likely to be surrounded by  a thick and
expanding radioactive ejecta. In the case  of GW170817, the optical depth is expected to be dominated by the r-process radioactively  powered, kilonova ejecta \citep[e.g.][]{2010MNRAS.406.2650M,2011ApJ...736L..21R,2017Natur.551...80K}. Given the  quantities derived  for the neutron star merger outflow of GW170817 \citep[e.g.][]{2017ApJ...851L..21V,2017Natur.551...80K, 2017ApJ...848L..34M,2017Sci...358.1583K,2019ApJ...883L...6R}, we expect the ejecta to become transparent after a time  
\begin{equation}
t_{\tau=1} \approx  1.1 \left({E_{\rm ej} \over  10^{50}{\rm erg}}\right)^{-1/2} \left({M_{\rm ej}\over 10^{-2}\, \mathrm{M}_\odot}\right) \;{\rm days},
\end{equation}
where we have used $\kappa \approx 10$~cm$^2$~g$^{-1}$ for the much more opaque r-process rich  ejecta \citep{2013ApJ...775...18B}. Given the  low mass ejecta, 
double neutron star  mergers appear to be a viable  system for uncovering a 
neo-neutron star provided that the surviving remnant is stable. 
Alternatively, the lack of X-ray detection of the cooling signal could be used to argue in support of a collapse to a black hole.

\subsection{Neo-Neutron Stars in Accretion Induced Collapse Events}

 Another relevant progenitor avenue for our study  is  the formation of a neutron star through the collapse of oxygen-neon white dwarf stars in interacting binaries \citep{1976A&A....46..229C,1980PASJ...32..303M, 1990ApJ...356L..51C, 1991ApJ...367L..19N, 2018RAA....18...49W}.  A oxygen-neon white dwarf  in a binary system might be able to augment  its mass near the Chandrasekhar mass leading to accretion-induced collapse by accreting steadily or dynamically \citep[e.g.][]{2019MNRAS.484..698R, 2018MNRAS.481..439W}.  The formation of neutron stars from interacting oxygen-neon white dwarfs  in binaries is likely to be accompanied by   low mass ejecta \citep{1992ApJ...391..228W, 2006ApJ...644.1063D, 2009MNRAS.396.1659M,2010MNRAS.409..846D}, which might help  direct detection.
 
The prospects for detection of the predicted  transients  appear promising \citep{2010MNRAS.409..846D}, yet their  characterization  might be difficult as they might be confused with other thermal transients predicted to occur on similar timescales ($\approx$ few days) such as failed deflagrations \citep{2005ApJ...632..443L} and  
type  .Ia supernovae   \citep{2007ApJ...662L..95B}. 
Such events are, however, not expected to be accompanied by an  X-ray transient. For an ejecta mass of $M_{\rm ej}= 10^{-2}\, \mathrm{M}_\odot$, we expect these optical transients to be uncovered  by upcoming surveys to  distances of a few 100 Mpc \citep{2010MNRAS.409..846D}, which will make the X-ray characterization of the neo-neutron star doable with current  space-based facilities. 

\section{Summary of Results and Conclusions}
\label{sec:Concl}

We have presented a detailed study of the evolution of the outer layers of a neo-neutron star.
We started just after the end of the proto-neutron star phase when the internal temperature has dropped to $\sim 2.5\times 10^{10}$ K
at densities $\sim 10^{11} \, \gcc$ and above.
At these temperatures the nuclei in the crust have already been formed.
We developed model of the outer envelope, i.e. the region from the photosphere up to densities around $\sim 10^{5-6} \, \gcc$,
at luminosities close to the Eddington luminosity,  in stationary state, presented in \Fig{fig:Env}.
Using an extension of the neutron star cooling code \texttt{NSCool} 
\citep{Page:1989kx,Page:2016fk} we then modeled the whole neutron star,
but focused on the description of the evolution of the inner envelope, at densities between $10^5$ to $10^{11} \, \gcc$.
The evolution of the surface temperature, and hence the star's surface thermal luminosity, during this early neo-neutron star phase in controlled by the evolution of the inner envelope and is thermally decoupled from the deeper layers on such short time scales.
The initial condition of temperature $\sim 2.5\times 10^{10}$ K at high densities and surface Eddington luminosity leaves some, 
but not much, space for variability of the temperature and luminosity in the inner envelope as shown in \Fig{fig:Initial}.
As a result, the surface luminosity remains close to the Eddington value, i.e., above $10^{38}$~erg~s$^{-1}$, for a few thousand seconds with 
effective temperatures of the order of $1.5-2 \times 10^7$ K, as presented in \Fig{fig:Cool1} for a 1.4 $\Msun$ star.
After $\sim 10^3$ seconds the surface temperature evolution is controlled by neutrino emission, initially
by pair annihilation  in the inner envelope for some $10^5$ seconds followed by plasmon decay 
until it has decreased to a few millions K and reached the ``early plateau'', well known from isolated neutron star cooling theory.
Models with either larger or lower surface gravity have an initially different Eddington luminosity but later follow a very similar cooling
trajectory during their first year of evolution, as illustrated in \Fig{fig:Cool2}.
At ages between $10^4$ and $10^7$ seconds the luminosity drop is roughly, within a factor of a few, a power law
\begin{equation}
\label{eq:Ldrop}
L(t) \simeq 3\times 10^{37} \left(t/10^4 \, \text{s}\right)^{-3/4} \;\mathrm{erg~s^{-1}} \; .
\end{equation}

Neutron stars in the universe could have very different origins, including core-collapse supernovae, neutron star mergers, white dwarf collapses,
and the so-called electron-capture supernovae that are somewhat similar to the accretion induced collapse.
In the case of birth in a core collapse supernova it is very unlikely that the neo-neutron star could be observed since it takes at least a few months
till the remnant could become transparent to soft X-rays.
In the case of SN 1987A, it is only after more than 30 years that a first signal 
of the existence of a neutron star has been possibly observed
\citep{Cigan:2019aa}.
However, in the case of a born-again neo-neutron star produced by the merging of two neutron star, as is expected to be the case in short GRBs and
in the GW170817 event, chances of detection of the neo-neutron star are encouraging.
If the merger produces a GRB and we are strongly off-axis the neo-neutron star could be detectable once the kilonova ejecta have sufficiently expanded
to become transparent to soft X-rays, which should take about a day (see Equation 26).
In the case of accretion induced collapse of a white dwarf the situation is similar with little ejected material.

The direct detection  of a neo-neutron star can thus be aided if the formation is followed by the ejection of low mass ejecta as in the case of neutron star  mergers and accretion-induced collapses of white dwarfs.  Interestingly, in both scenarios the neo-neutron star  is expected to be rapidly  rotating and if it has a sufficiently high magnetic field, then the spin-down luminosity might  prevent the detection of the cooling signature
\citep{Rosswog:2003ab,Price:2006aa}.
Our understanding of neutron star birth  has come a long way since the pioneering work by \citet{1934PNAS...20..254B}
more than 8 decades  ago, but these enigmatic sources continue to remain elusive, in particular at very young ages.  
Neutron star  mergers and accretion-induced collapses of white dwarfs provide us with an exciting opportunity to study new regimes of physics 
and to learn what these systems  were  like at the earliest epochs of formation when their luminosities are near the Eddington limit.
Electromagnetic and gravitational-wave observatories over the coming years offer the potential to uncover the detailed nature of these most remarkable objects.

\acknowledgments
M.B. and D.P. are partially supported by the Consejo Nacional de Ciencia y Tecnolog{\'\i}a with a CB-2014-1 grant $\#$240512.
M.B. also acknowledge support from a postdoctoral fellowship from UNAM-DGAPA. 
The first steps of this work we financed by a UNAM-DGAPA grant \#113111.
E.R.-R. acknowledge support from the Heising-Simons Foundation,  the Danish National Research Foundation (DNRF132) and NSF (AST-1911206).
The authors thank A. Raduta for some guidance about the nuclei specific heat and J. Schwab, R. Foley  and L. Lopez for illuminating discussions.

\rev{\software{NSCool \citep{Page:1989kx,Page:2016fk}}}\footnote{The 
version of the code used in the paper is not publicly available at the present 
time, but the modifications are described in \App{sec:Solver}}

\appendix

\section{Physical State of Matter in the Inner Envelope}
\label{sec:NeoNS:Phys}

As discussed in Sect.\ \ref{sec:NeoNS} in the inner envelope we are dealing with the matter at densities $\rhob=10^5 < \rho < \rhoc=10^{11}~\gcc$ and 
temperatures $(1-3) \times 10^9 < T < (1-3)\times 10^{10}$~K. 
At this temperatures and densities one has to take into account presence of positrons and photons.
Below we briefly describe the physical ingredients employed in our model and 
present illustrative plots that allow to better understand
our results.

\subsection{Equation of State}
\label{sec:NeoNS:EOS}

We assume the presence of $^{80}$Ni at $\rhoc$ (following \citealt*{Haensel:1989vn}) and $^{56}$Fe at $\rhob$, while we interpolate in both $A$ and $Z$ 
linearly in $\log \rho$ at intermediate densities.
Pressure is obtained as the sum of radiation, free gases of electrons and positrons, a free gas of nuclei plus Coulomb interaction corrections 
following \citet{PC10}.
Crystallization of ions takes place when the Coulomb coupling parameter $\Gamma \equiv \slfrac{(Ze)^2}{(a_\mathrm{WS} k_\mathrm{B}T)}$ reaches 175
[$a_\mathrm{WS} = (\slfrac{4\pi n_\mathrm{i}}{3})^{-1/3}$ is the Wigner-Seitz cell radius, $n_\mathrm{i}$ being the number density of ions and $Ze$ their electric charge].

\subsection{Opacity and Thermal Conductivity}
\label{sec:NeoNS:Conduct}

\begin{figure*}
	\gridline{\fig{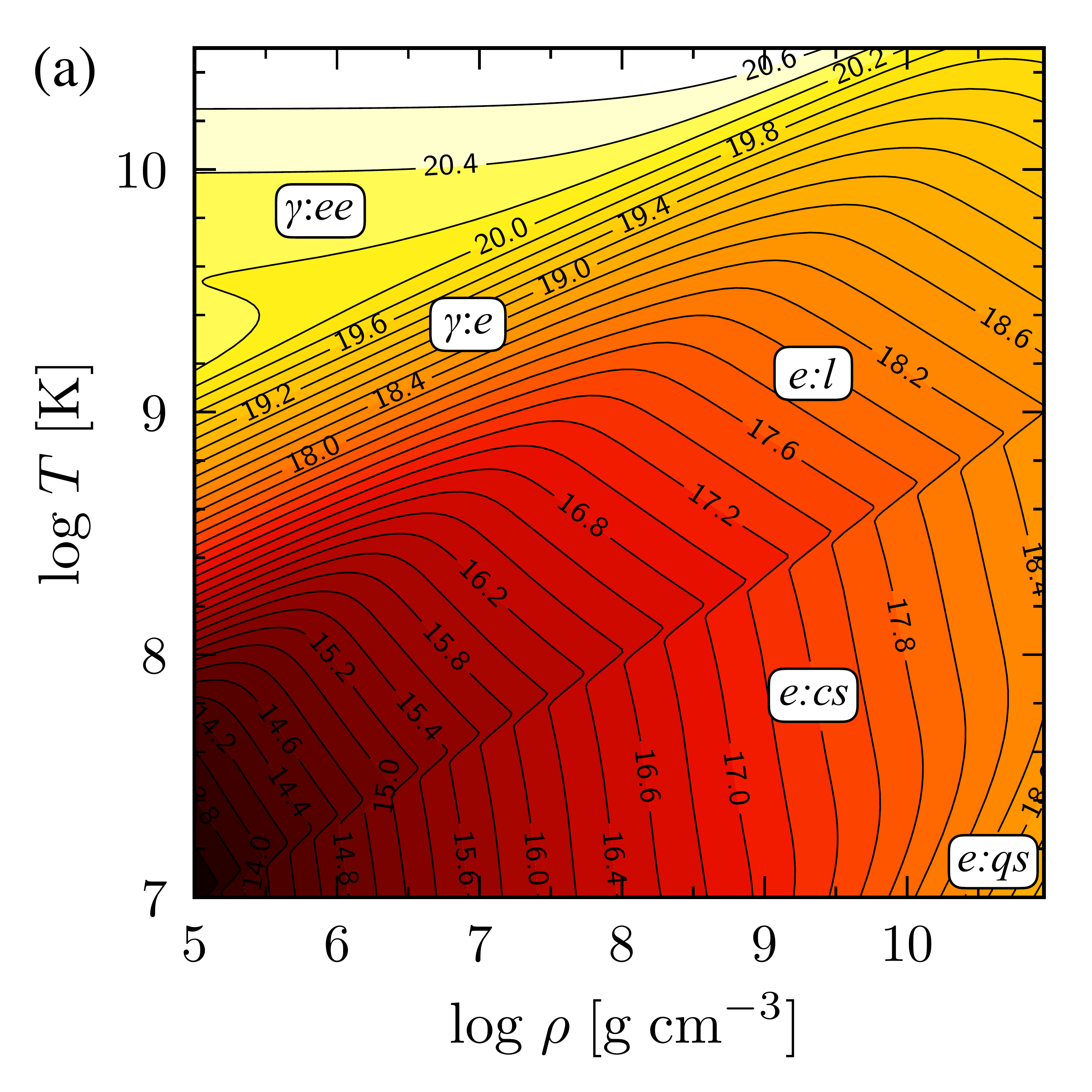}{0.46\textwidth}{}
 		     \fig{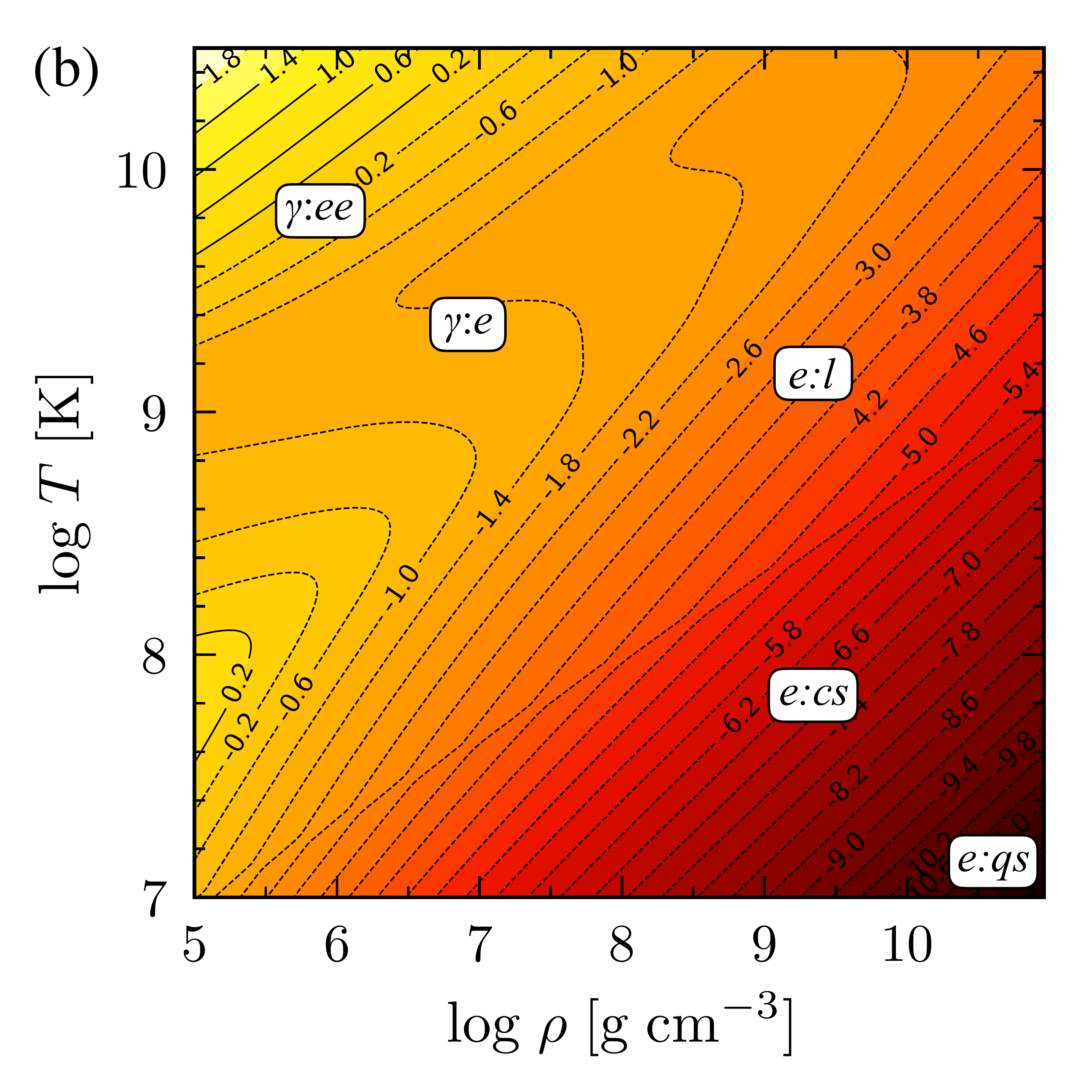}{0.46\textwidth}{}}
	\vspace{-0.3in}
	\gridline{\fig{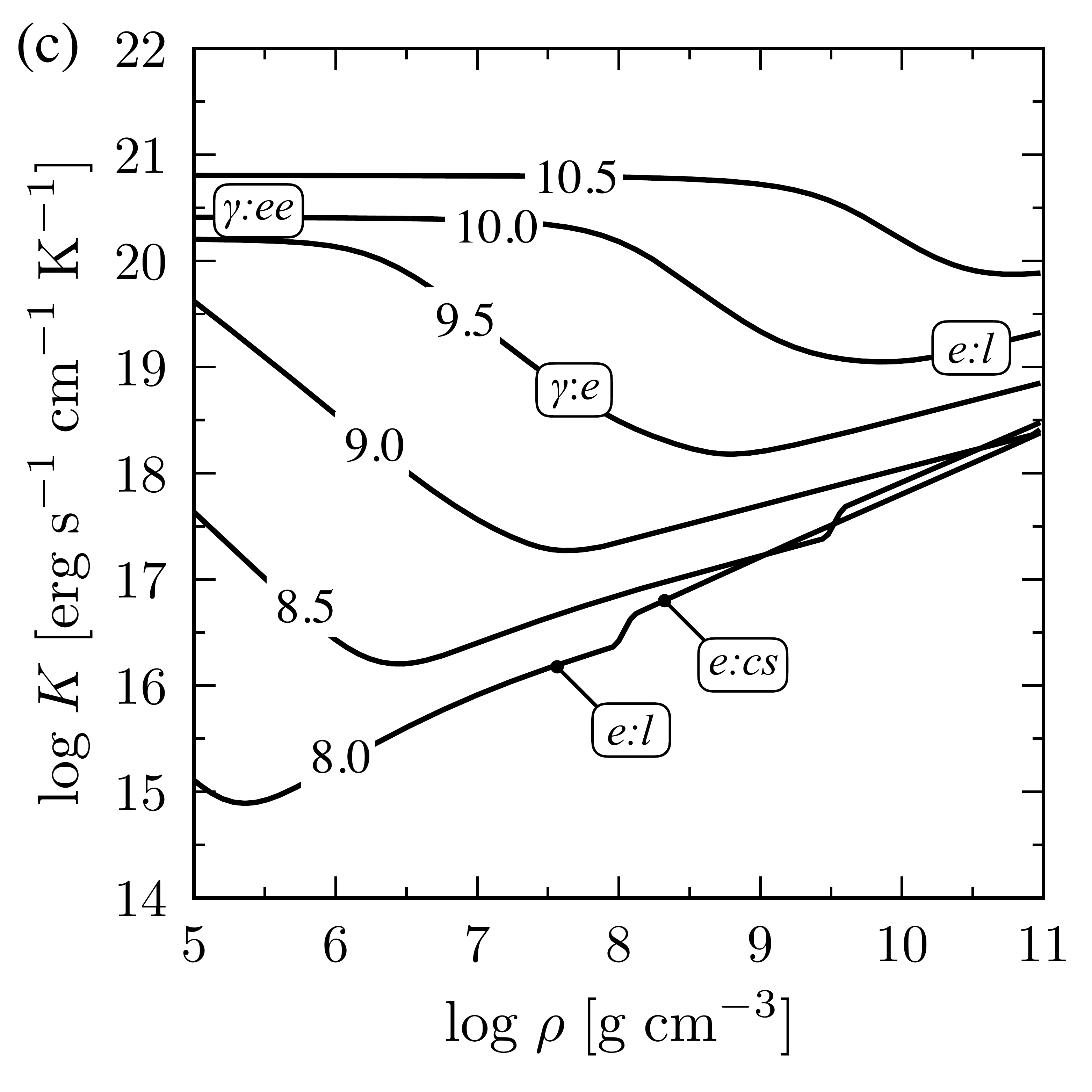}{0.46\textwidth}{}
		     \fig{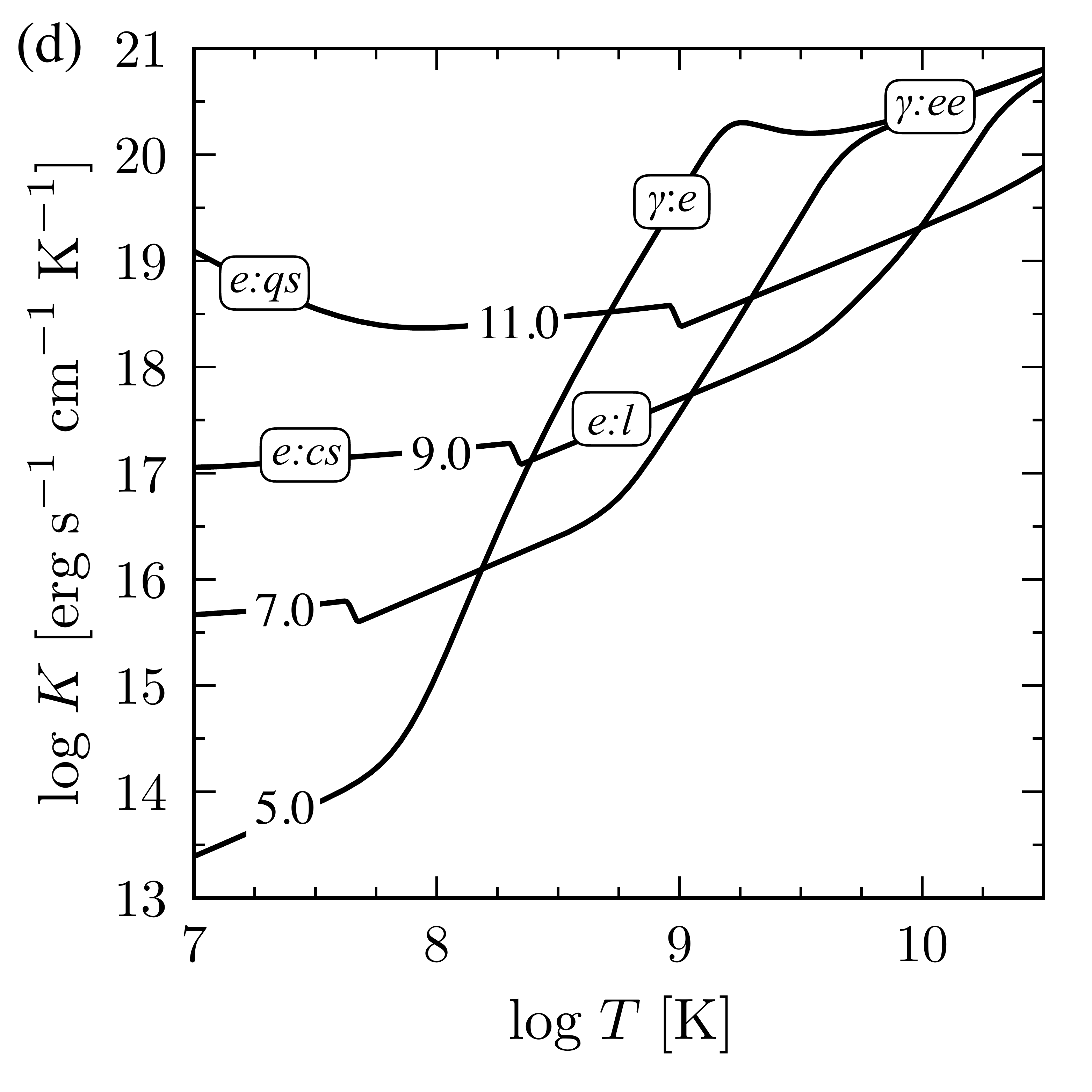}{0.46\textwidth}{}}
	\vspace{-0.3in}
	\caption{Thermal conductivity $K$ and total opacity $\kappa$ for the conditions of the inner envelope. 
	Panel (a) shows the contour plot of conductivity, panel (b) of opacity (from \Eq{Eq:Kkappa}), panels (c) and (d) demonstrate the thermal  conductivity as a function of density, for a constant temperature, and temperature, for a constant density, respectively.
	Values on the contour lines on panel (a) are decimal logarithms of the conductivity [in erg~s$^{-1}$~cm$^{-1}$~K$^{-1}$], on panel (b) decimals logarithms of opacity  [in cm$^2$~g$^{-1}$], on curves on panel (c) are decimal logarithms of temperature [in K] and values on curves on panel (d) are decimal logarithms of density [in $\gcc$]. The boxed labels in panels (a) and (b) indicate the dominant contribution to the thermal conductivity for a given temperature and density region, as described in the text, and a few of them are reproduced in the other two panels. 
	}
	\label{fig:Conduct}
\end{figure*}

The thermal conductivity $K$ is taken as the sum of the electron, $K_\mathrm{e}$, and photon $K_\mathrm{ph}$ conductivities.
In the inner envelope the plasma is fully ionized. Thus, there is no need to take into account the effects of partial ionization on opacity. The radiative opacity
$\kappa_\mathrm{rad}$ consists of two terms: free-free absorption and electron scattering. The former was calculated based on the fits of \citet{Schatz99}. The latter is
based on the modern fit of \citet{Poutanen17}, which takes into account electron degeneracy and pair production (the fit handles both Thompson and Compton
scattering).
 A correction factor of \citet{PY01} was used for adding free-free and electron-scattering opacities.
The electron thermal conductivity is taken from \citet{Yakovlev:1980aa} when ions are in a liquid phase and from
\citet{Potekhin:1999aa} in the solid phase.

The resulting thermal conductivity $K$ is illustrated in Fig.~\ref{fig:Conduct} as well as the corresponding total opacity $\kappa$ defined by
\begin{equation}
\kappa = \frac{4acT^3}{3 K \rho} \, .
\label{Eq:Kkappa}
\end{equation}
This $\kappa$, which include contributions from photons and electrons, should not be confused with the more restricted (Rosseland mean) radiative opacity.

The different shape of the contour lines in panel (a) clearly exhibit different regimes that are indicated by the boxed labels:
\begin{itemize}[nosep]
	\item $\gamma\!:\!ee$ -- conductivity dominated by photons and controlled by Thomson/Compton scattering on electrons and positrons; 
	\item $\gamma\!:\!e$ -- conductivity dominated by photons and controlled by Thomson scattering on electrons;
	\item $e\!:\!l$ -- conductivity dominated by electrons and controlled by scattering on ions in the liquid phase;
	\item $e\!:\!cs$ -- conductivity dominated by electrons and controlled by scattering on ions in a classical Coulomb solid;
	\item $e\!:\!qs$ -- conductivity dominated by electrons and controlled by scattering on ions in a quantum Coulomb solid;
\end{itemize} 
In the density regime considered here for the inner envelope, photon opacity is dominated by free-free absorption only in a very narrow region
at the transition between photon dominated to electron dominated transport.
We have a discontinuity in the electron conductivity along the melting curve which may be fictitious as argued by \citet{Baiko:1998aa}
but which is small enough and occurring at densities high enough that it has a negligible effect on our results.
Notice the dramatic effect of pairs in limiting $K$ at high temperatures simply due to the strong increase of the number of scatterers,
electrons and positions, with temperature in this regime.

\begin{figure}
	\begin{minipage}[t]{0.47\textwidth}
		\includegraphics[height=8.0cm,keepaspectratio=true,clip=true,trim=0.45cm
		 0.5cm 0.7cm 0.6cm]{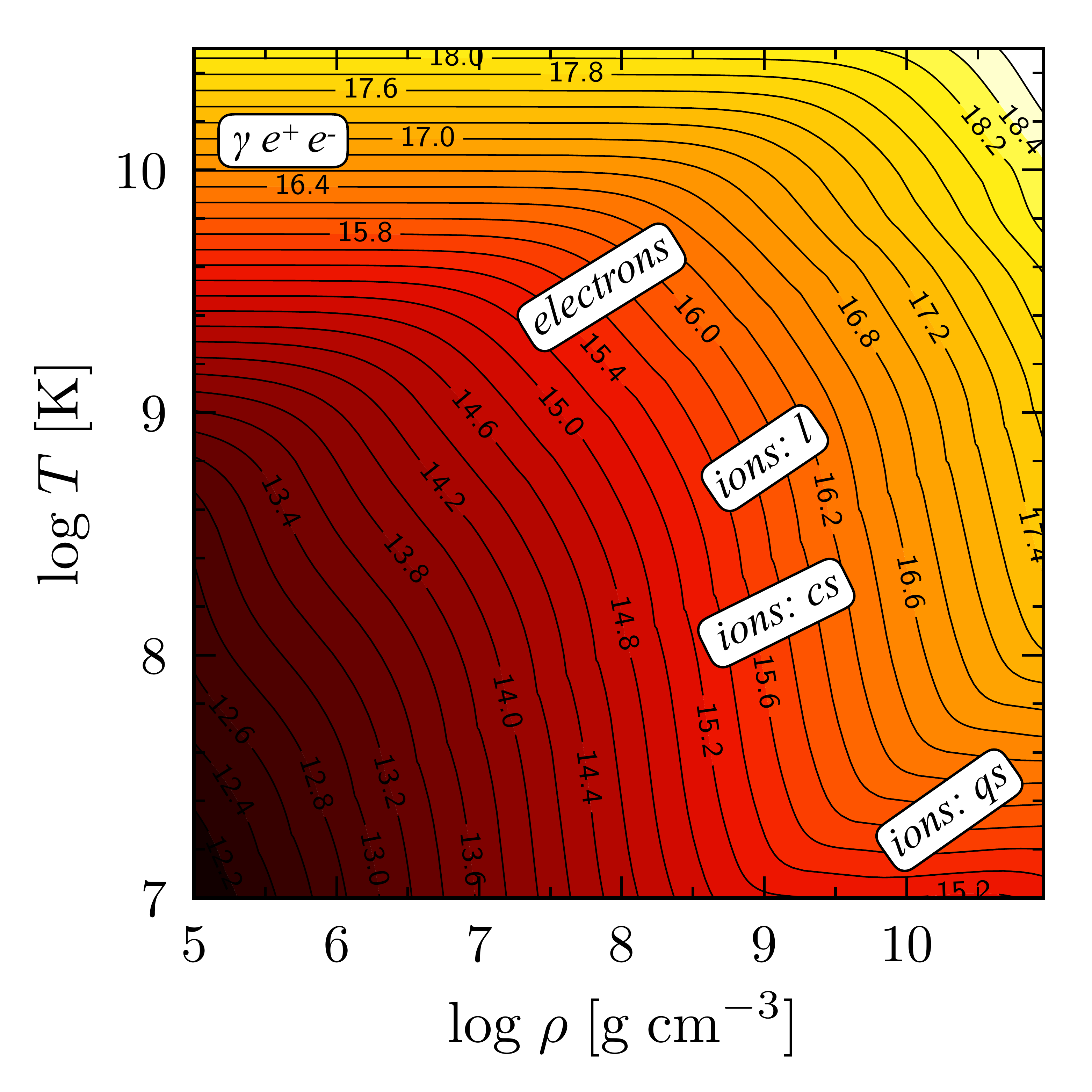}
		\caption{Specific heat  for the conditions of the inner envelope. 
		Values on the contour lines are decimal logarithms of the specific heat 
		capacity [in erg~cm$^{-3}$~K$^{-1}$].
		Boxed labels indicate the dominant contributor in the various temperature 
		and density regions. See details in the text.}
		\label{fig:Cv_2D}
	\end{minipage} \hfill
	\begin{minipage}[t]{0.47\textwidth}
%
		\includegraphics[height=8.0cm,keepaspectratio=true,clip=true,trim=0.45cm
		 0.5cm 0.7cm 0.6cm]{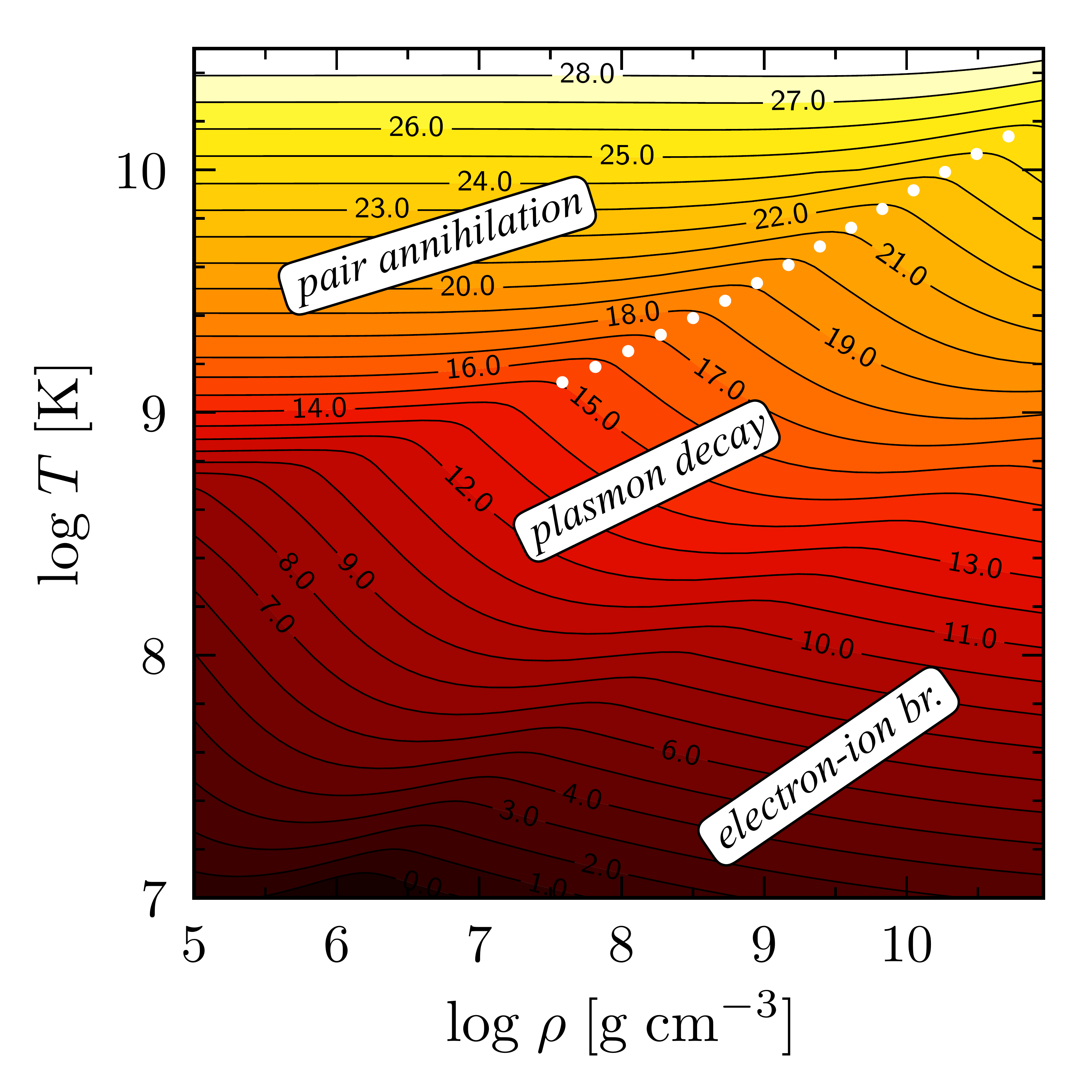}
		\caption{Neutrino emissivity for the conditions of the inner envelope. 
		Values on the contour lines are decimal logarithms of the neutrino 
		emissivity [in erg~cm$^{-3}$~s$^{-1}$].
		Boxed labels indicate the dominant process in the various temperature and 
		density regions
		(where ``electron-ion br.'' stands for electron-ion bremsstrahlung) and the 
		thick dotted (white) line
		explicitly marks the transition form pair annihilation to plasmon decay for 
		later reference
		(the line is not shown at low densities because neutrino losses become 
		negligible in this regime).
		See details in the text.}
		\label{fig:Nu_2D}
	\end{minipage}
\end{figure}
	
\subsection{Specific Heat}
\label{sec:NeoNS:Cv}

\begin{figure*}
	\includegraphics[height=8.0cm,keepaspectratio=true,clip=true,trim=0.45cm 0.5cm 0.7cm 0.6cm]{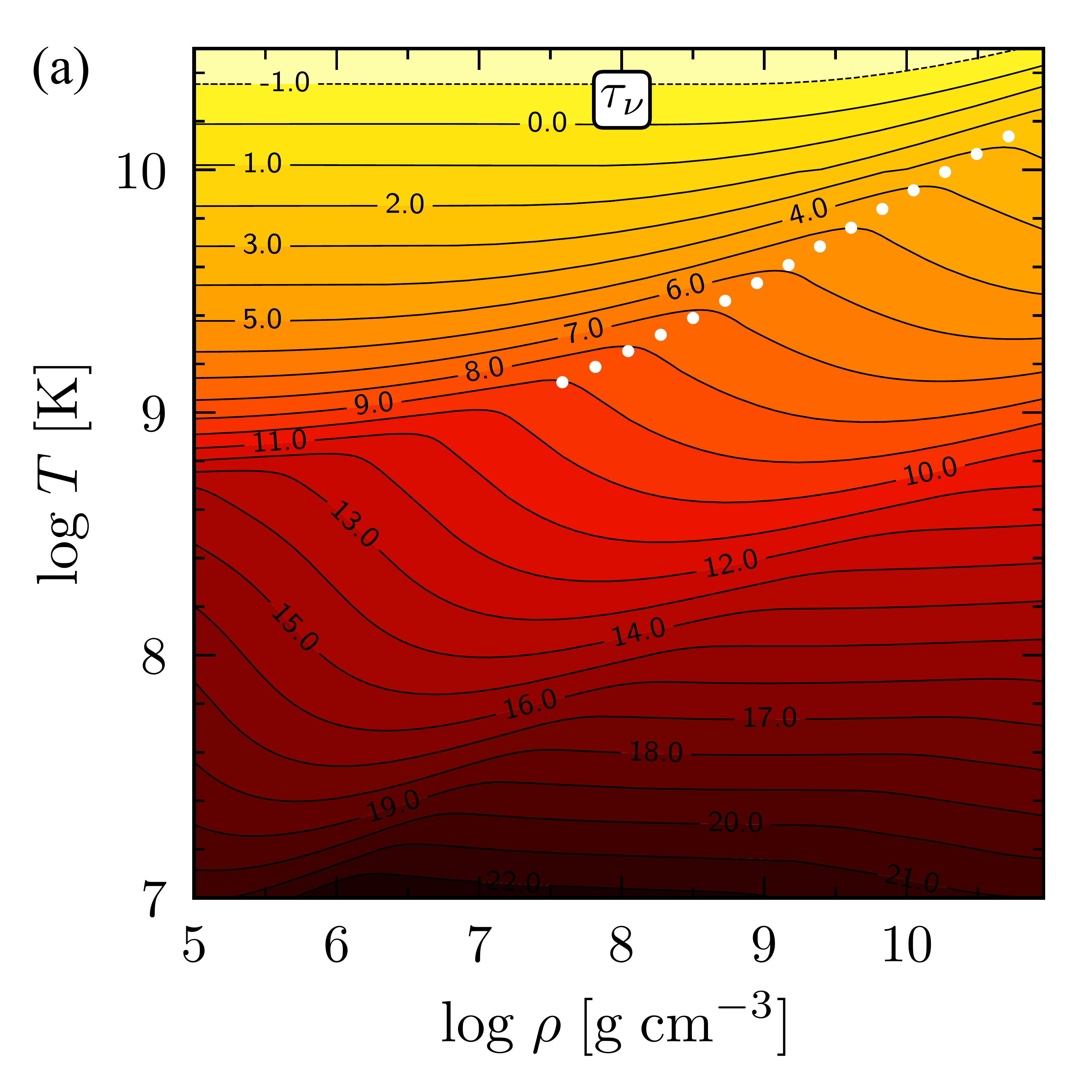}
	\hspace{1.5cm}
	\includegraphics[height=8.0cm,keepaspectratio=true,clip=true,trim=0.45cm 0.5cm 0.7cm 0.6cm]{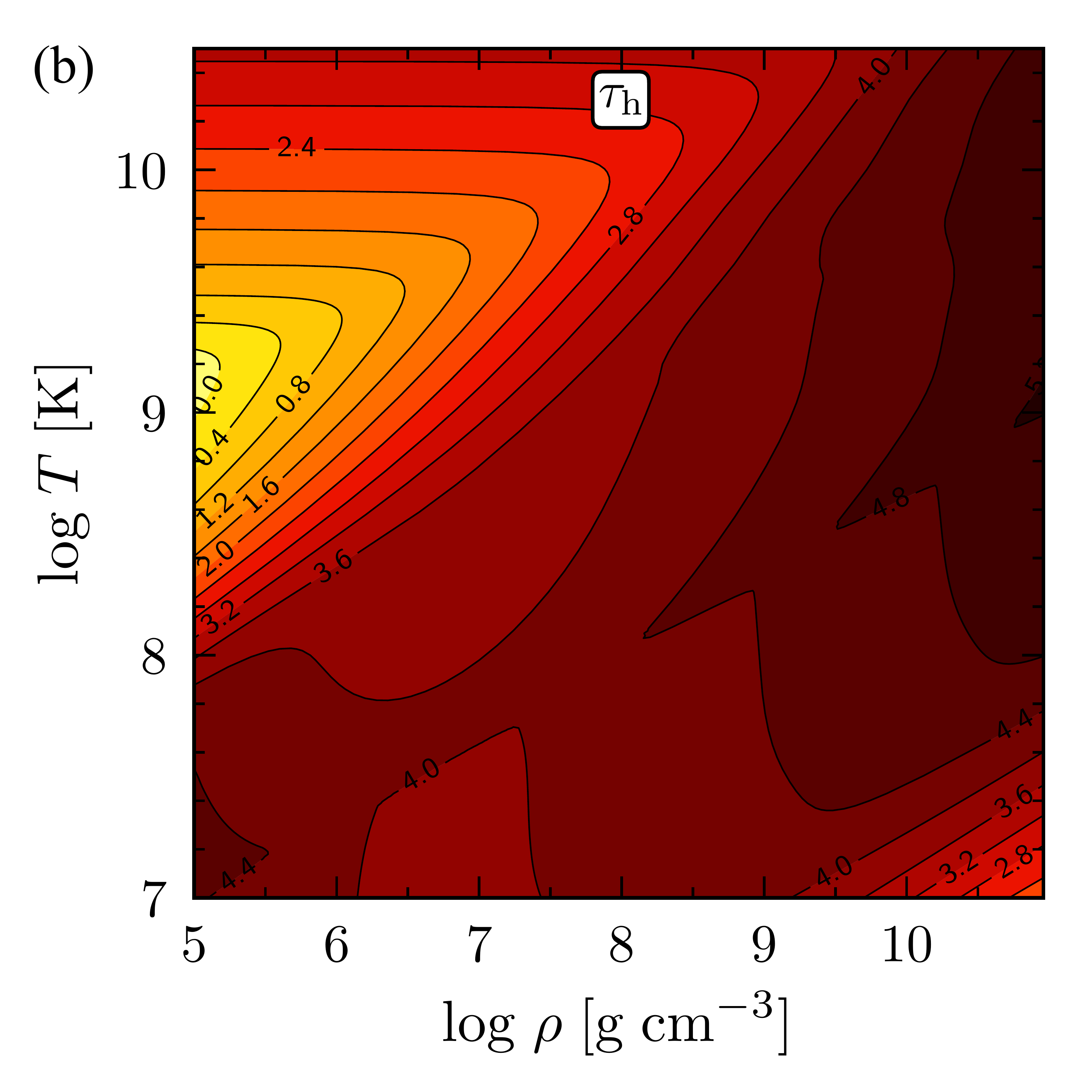}
	\caption{Time-scales in the inner envelope.
	Panel (a) shows contours of the neutrino cooling time scale $\tau_\nu \equiv C_V/Q_\nu$, the curve labels giving the decimal logarithms of 
	$\tau_\nu$ in units of second per GK (since $10^9$ K is a typical temperature in our neo-neutron star envelopes).
	The thick dotted (white) line reproduces the one from \protect\Fig{fig:Nu_2D}.
	Panel (b) shows contours of the heat diffusion time scale $\tau_\mathrm{h} \equiv C_V/K$, the curve labels giving the decimal logarithms  of
	 $\tau_\mathrm{h}$ in units of second per (10 meters)$^2$ (since 10 meters is a typical length-scale in our neo-neutron star envelopes).
	}
	\label{fig:Scales}
\end{figure*}

The specific heat is computed as described in \citet{PC10} to which we added the contribution of radiation and pairs when present.
As we are interested in temperatures up to about $1-2$~MeV we have moreover added the contribution of nuclear excitations
as discussed in \App{app:NCv}.

Contour plots of the total specific heat are shown on Fig.~\ref{fig:Cv_2D}. 
The different regimes, as indicated by the boxed labels are:
\begin{itemize}[nosep]
	\item $\gamma e^+ e^-$ -- photons and electron-positron pairs;
	\item $\mathit{electrons}$ -- electrons;
	\item $\mathit{ions: l}$ -- ions in a Coulomb liquid;
	\item $\mathit{ions: cs}$ -- ions in a classical Coulomb solid;
	\item $\mathit{ions: qs}$ -- ions in a quantum Coulomb solid.
\end{itemize} 
Nuclear excitation never dominate but make a significant contribution at the highest densities ($\geq 10^{10}~\gcc$) and temperatures ($\gg 10^9$ K).

\subsection{Neutrino Emission}
\label{sec:NeoNS:Nu}

In the density range of our inner envelope, $10^5 - 10^{11} \gcc$, neutrino emission is dominated by three process, in order of decreasing temperature
importance: $e^+-e^-$-pair annihilation, plasmon decay and e-ion bremsstrahlung.
For the first two we follow \citet{Itoh:1996aa} and \citet{Kaminker:1999aa} for the third one.
We present in \Fig{fig:Nu_2D} contour plots of the total neutrino emissivity. 
Notice the dramatic change in temperature dependence when crossing the (dotted white) line from pair annihilation to plasmon decay dominance.
The very strong temperature dependence of the pair annihilation process when approaching this line is due to the exponential suppression of pairs
when electrons become degenerate.
Similarly, when shifting from plasmon decay to electron-ion bremsstrahlung the temperature dependence of the plasmon process increase
rapidly due to the exponential suppression of plasmons below the plasmon temperature.

\subsection{Time-Scales}
\label{sec:NeoNS:times}

Besides the micro-physics ingredients, $\kappa$, $C_V$, and $Q_\nu$, the two evolutionary time-scales dictated by them are also very illustrative:
the neutrino cooling time scale $\tau_\nu \equiv C_V/Q_\nu$ and the heat diffusion time scale $\tau_\mathrm{h} \equiv C_V/K$.
We display both of them in \Fig{fig:Scales} as they are very helpfull to understand our results.

\section{Nuclei specific heat}
\label{app:NCv}

In principle, calculation of the specific heat is straightforward as it can be directly derived from the system's partition function $Z = Z(T)$ (see, e.g., \citealt{LL_Stat93}) as
\begin{equation}
	C_V = k_\mathrm{B} T \left(2 \frac{Z'}{Z} - T \left( \frac{Z'}{Z}\right)^2 + T \frac{Z''}{Z} \right),
\label{Eq:A:Cv}
\end{equation}
where primes denote derivative with respect to the temperature $T$ and $k_\mathrm{B}$ is the Boltzman constant. 
We describe below how do we proceed to calculate the partition function. 

The nucleus has a discrete excitation energy spectrum but only low lying energy levels (up to a few MeV) are known reliably from experiments. 
Moreover, at higher energies the density of states grows so rapidly that it is more convenient to approximate with a continuous distribution.
If $\rho_\mathrm{LD}(E,J)$ is the density of energy levels of angular momentum $J$ at energy $E$, the ``observable level density''
is $\rho_\mathrm{LD}(E) = \sum_J \rho_\mathrm{LD}(E,J)$ while the density of sates or ``true level density'', that takes into account the spin degeneracy, is
$ \Omega_\mathrm{DS}(E) = \sum_J (2J+1) \,\rho_\mathrm{LD}(E,J)$ \citep{GC65,Huizenga:1972aa}.
We will follow the commonly used  Back-Shifted Fermi Gas (BSFG) approximation (see, e.g., \citealt{EB05}), 
an extension of the non-interactive Fermi gas model of \citet{Bethe36}, in which
\begin{equation}
	\Omega_\mathrm{DS}(E) =  \frac{\sqrt{\pi} \exp{\left(2\sqrt{a(E-E_1)}\right)}}{12\, a^{\slfrac{1}{4}} (E-E_1)^{\slfrac{5}{4}}} = 
	\sqrt{2\pi} \sigma \, \rho_\mathrm{LD}(E)
\label{Eq:A:DS}
\end{equation}
where $E_1$ is the energy back-shift, $a$ the level density parameter, and $\sigma$ the spin cutoff, whose values are obtained by fitting experimental data.
Within this approximation we can calculate $Z(T)$ as
\begin{equation}
	Z(T) =  \sum_{i=0}^{i=i_\mathrm{cf}} g_i \exp{ \left( -\frac{E_i}{k_\mathrm{B} T} \right)} + 
	          \int_{E_\mathrm{cf}}^{\infty} \Omega_\mathrm{DS}(E) \exp{ \left( -\frac{E}{k_\mathrm{B} T} \right)} \dd E \quad ,
\label{Eq:A:PF}
\end{equation}
where $E_\mathrm{cf} = E_{i_\mathrm{cf}}$ is some arbitrary energy cutoff level to switch from discrete to continuous regime; $g_i$ is the spin degeneracy factor, which is obtained experimentally together with the energy levels $E_i$.
In the continuous spectrum range the spin degeneracy is in principle taken into account in $\Omega_\mathrm{DS}(E)$.
This spin degeneracy is not experimentally determined in the high excitation (continuous) regime and there are large uncertainties in its value
(see, e.g., \citealt{von-Egidy:2009aa}) and for this reason many authors prefer to use $\rho_\mathrm{LD}(E)$ instead of $\Omega_\mathrm{DS}(E)$
in the evaluation of $Z(T)$ in \Eq{Eq:A:PF}.

\begin{figure*}
	\centering
	\includegraphics[height=8.0cm,keepaspectratio=true,clip=true,trim=0.1cm 0.5cm 0.7cm 0.1cm]{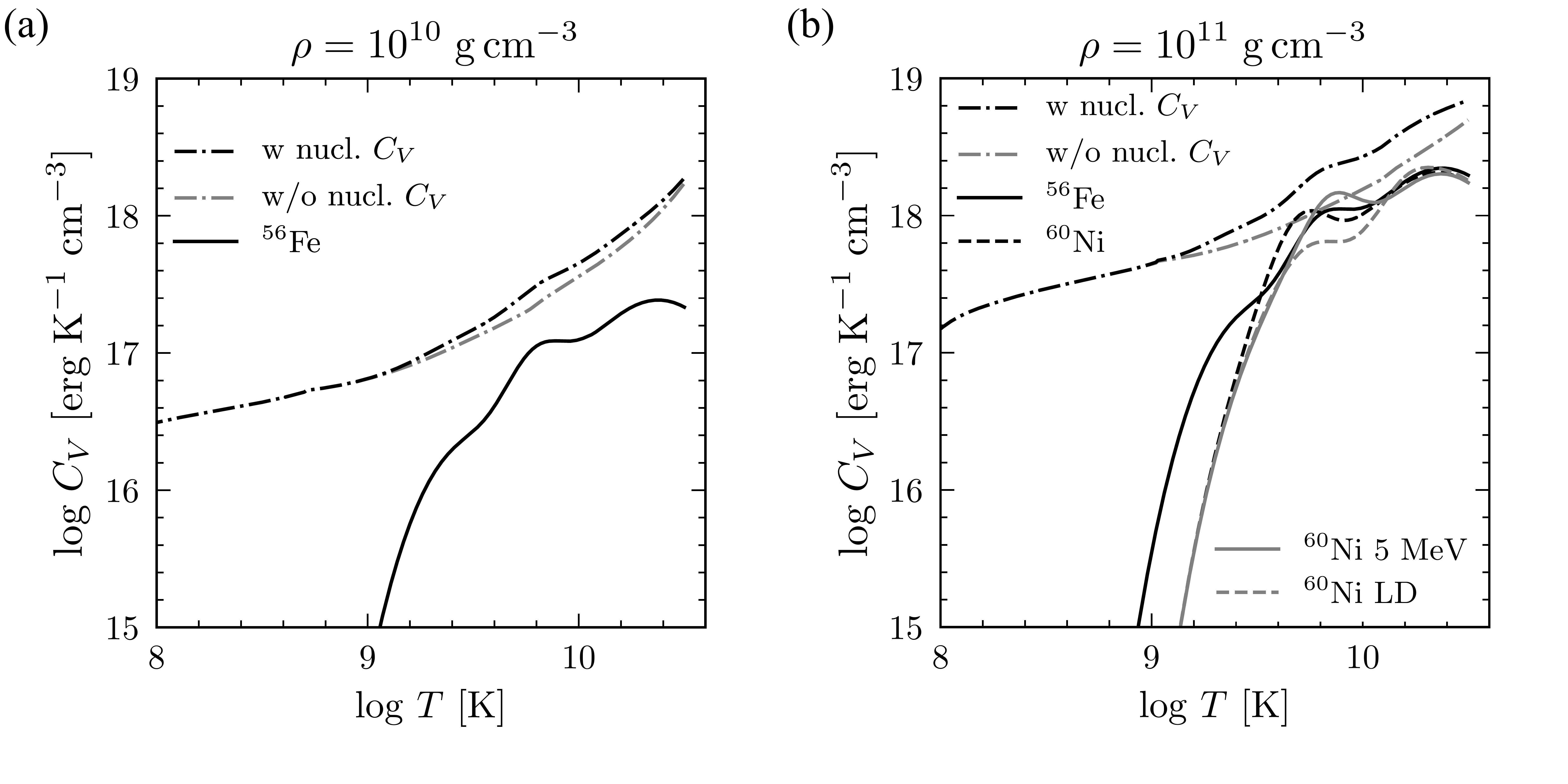}
	\caption{Heat capacity, of both the nuclei and the total, as a function of temperature at two fixed densities $\rho = 10^{10}$, panel (a), and $10^{11}~\gcc$, panel (b).
	Solid black curve corresponds to $^{56}$Fe, dashed black curve -- to $^{60}$Ni, dot-dashed black curve -- to total heat capacity including contribution from $^{56}$Fe nuclei, dot-dashed gray curve -- to total heat capacity excluding nuclei contribution. Solid gray curve demonstrates  $^{60}$Ni nuclei heat capacity for the energy cutoff of  5~MeV. Dashed gray curve shows $^{60}$Ni nuclei heat capacity calculated employing the level density  $\rho_\mathrm{LD}$  instead of the density of states $\Omega_\mathrm{DS}$. See details in the text.}
	\label{fig:A:NCv}
\end{figure*}

We used the procedure described above to calculate the contributions of $^{56}$Fe and $^{60}$Ni nuclei to the heat capacity. For $^{56}$Fe we used experimental data on the energy levels and spin degeneracy factors from \citet{NDS_Fe56} and for $^{60}$Ni -- from \citet{NDS_Ni60}.\footnote{All these data are available online at \url{http://www.nndc.bnl.gov/chart/}} The values of $E_1$ and $a$ were taken from the recent fits of \citet{BE15}. The choice of the cutoff energy is important, so we tried two approaches: cutoff at 5 MeV and cutoff at the first energy level for which experimental value of $g$ is not known. In the latter case the cutoff energy was $\sim 3.8$~MeV for $^{56}$Fe and $\sim 3.3$~MeV for $^{60}$Ni. The heat capacity calculated using Eqs. \eqref{Eq:A:Cv}, \eqref{Eq:A:DS} and \eqref{Eq:A:PF} is per nucleus. Thus, one have to multiply it by the ion number density $n_\mathrm{i} = \slfrac{\rho}{(A m_u)}$, where $A$ is the nucleus atomic mass number and $m_u$ is atomic mass unit.
The results are presented on Fig.~\ref{fig:A:NCv}, which shows the heat capacity, of both the nuclei and the total, as a function of temperature at two fixed densities $\rho = 10^{10}$ and $10^{11}~\gcc$.  
Solid black curve corresponds to $^{56}$Fe, dashed black curve -- to $^{60}$Ni, dot-dashed black curve -- to total heat capacity including contribution from $^{56}$Fe nuclei, dot-dashed gray curve -- to total heat capacity excluding nuclei contribution. Solid gray curve demonstrates  $^{60}$Ni nuclei heat capacity for the energy cutoff of  5~MeV instead of $\sim 3.3$~MeV (see previous paragraph). Dashed gray curve shows $^{60}$Ni nuclei heat capacity calculated employing the level density  $\rho_\mathrm{LD}$  instead of the density of states $\Omega_\mathrm{DS}$  in \Eq{Eq:A:PF}.

From Fig.~\ref{fig:A:NCv} one can make several conclusions. First, at low temperatures ($T \sim 10^9$~K) $^{56}$Fe and $^{60}$Ni heat capacities are considerably different. This is not a surprise because at these temperatures the heat capacity is governed by a first few low lying energy levels, which can differ rather noticeably even for similar nuclei. On the other hand, this does not matter much as at $T \sim 10^9$~K nuclei heat capacity is much less than the total heat capacity and, thus, can be neglected. Second, the maximum contribution of the nuclei heat capacity to the total heat capacity is achieved at $T \sim 10^{9.8}$~K and can be around 50\% of the total heat capacity at $\rho = 10^{11}~\gcc$. Nuclei heat capacity is directly proportional to the density of matter (see paragraph before previous), thus, its contribution at lower densities is lower and at sufficiently low densities ($\rho \lesssim 10^{9}~\gcc$) can be neglected at any temperature. 
Third, at temperatures $T \sim 10^{9.6} - 10^{10}$~K nuclei heat capacity is sensitive to the particular nuclear species and to the choice of the energy cutoff. The difference can be up to $\sim 2$ times. 
At higher densities one enters the neutron drip regime where the heat capacity is dominated by the dripped neutron liquid 
(see e.g., \citealt{Page:2012ys}) and the contribution from the nuclear excitations can again be neglected. 
%

\section{Numerical Method}
\label{sec:Solver}

We base our calculations on the code \texttt{NSCool} 
\citep{Page:1989kx,Page:2016fk} with important adjustments to solve for 
hydrostatic equilibrium in the inner envelope
in the conditions where radiation and pairs pressures are important.

The structure equations are initially solved from the center of the star down to 
$\rhoc = 10^{11}~\gcc$ 
employing the zero temperature EOS and this interior structure is not modified afterward.
At densities between $\rhoc$ and $\rhob$ the structure equations are solved at 
every time step.
The thermal evolution equations (\ref{Eq:Thermal_T}) and (\ref{Eq:Thermal_L}) are solved in the whole star, i.e., from the center down to $\rhob$, at every time step.

So, in the inner envelope structure and thermal evolution equations have to be 
solved at each time step. There are several ways to do it and we had tried some 
of them until we have found a suitable one. The most considerable difficulty lies 
in the fact that the outer parts of the inner envelope are dominated by photons 
and electron-positron pairs. Thus, the adiabatic index is close to 4/3 and the 
system is close to being unstable.  

In the standard long term cooling calculation scheme (see, e.g., 
\citealt{Page:1989kx,Gnedin:2001aa}) thermal evolution 
equations are usually solved fully implicitly employing Newton-Raphson method 
(Henyey scheme \citealt{Henyey:1959uq}). The easiest way to modify this 
scheme to handle neo-neutron stars is to solve structure equations 
\emph{separately} from thermal equations at each Newton-Raphson iteration 
for the thermal equations. Unfortunately, this idea does not work. The thermal 
equation \eqref{Eq:Thermal_L} [which is basically the energy conservation law] 
and the hydrostatic equilibrium equation [Eq.~(4) of \citet{Potekhin:2018jb}] 
have a tendency to create oscillations in pressure, radius and temperature.
This is easy to understand: if we solve them \emph{separately}, some decrease 
in the radius will cause an increase in the temperature due to the injection of
contraction energy [Eq.~\eqref{Eq:Thermal_L}], which will increase the 
pressure and cause an increase in the radius due to the hydrostatic equilibrium 
equation. This will, in turn, cause the temperature and pressure to 
drop and a decrease in the radius. Clearly, this method is prone to instability and 
should not be used. We implemented it and found out that it indeed resulted in 
diverging iterations and in oscillations.   

So, to deal with this tendency to oscillate one has to solve structure and thermal 
equations \emph{together} in a single Newton-Raphson iteration scheme. In this 
case the changes in the pressure, radius and temperature are coordinated with 
each other at each iteration and consistent solution can be obtained. As it turns 
out there is now need to solve all six equations together in a single 
Newton-Raphson scheme. Actually, it is sufficient to solve only four equations 
[Eqs.~\eqref{Eq:Thermal_T}, \eqref{Eq:Thermal_L} and (1), (4) from 
\citet{Potekhin:2018jb}]  together and equations (2) and (3) from 
\citet{Potekhin:2018jb} can be solved separately as they do not 
produce any oscillations. We implemented this approach and it worked. 
However, the iterations converged slowly and not for all initial conditions.

So, we improved our solver further and implemented ``globally convergent'' Newton scheme of \citet{NR3}, which employed  backtracking line searches. It improved the situation. The iterations converged faster. But still the number of iterations for early timesteps was 5--10 times bigger than in the standard long term cooling. Probably, some even more elaborate solver can improve the situation yet it looks like that not much more. The alternative is to switch from a sequence of hydrostatic equilibriums to a hydrodynamic calculation in full GR, but this is far beyond the scope of current paper.

Another numerical complication is the necessity to match the initial and boundary conditions. For the standard long term cooling it is possible to start with constant (red-shifted) temperature profile (thus, zero luminosity inside, which is inconsistent with non-zero surface luminosity, i.e. $L_\mathrm{s} \neq L_\mathrm{b}$) and the matching will occur automatically at the first time step. 
In our neo-neutron stars models the Henyey method would not converge at the first time step if such inconsistent surface luminosity is employed.
So, we have developed a special matching procedure for the luminosity to start with the consistent initial and boundary conditions: $T_{t=0}(\rho) = F\big(\rho, \{p_1,p_2,\ldots \},p_\mathrm{match}\big)$, where $p_1,p_2,\ldots, p_\mathrm{match}$ are free parameters of the parametrization of an arbitrary initial temperature profile. The procedure is as follows: we fix the values of $p_1,p_2,\ldots$ and use Newton-Raphson method to search for the value of $p_\mathrm{match}$ until the initial profile satisfies the boundary condition to the desired precision [i.e, we stop when $L_\mathrm{s}\{T_\mathrm{s}(\Tb)\} = L_\mathrm{b}$]. Typically, this takes 5-6 Newton-Raphson iterations. Employment of such a procedure means that our initial temperature profile is no longer completely arbitrary.	

In particular, as an initial temperature we take a uniform value $T_0$ at densities above $\rhoc$ and in the inner envelope we choose 
\begin{equation}
	T_0(\rho) = T_{\mathrm{c},0}  - \Delta T \left(\frac{\log[\rhoc/\rho]}{\log[\rhoc/\rhob]} \right)^\gamma
\label{Eq:T_init}
\end{equation}
where $\Delta T = T_{\mathrm{c},0} - \Tb$, and $\gamma > 0$ is a power law index. So, if $\gamma$ is 1, then $T_0(\rho)$ is just linear in $\log{\rho}$. We usually fixed the value of $T_{\mathrm{c},0}$ to be $2.5 \times 10^{10}$~K and for various values of $\gamma$ we solved for $\Delta T$ to match the initial and boundary conditions.

Unfortunately, with the parametrization \eqref{Eq:T_init} the matching occurs 
only at super-Eddington surface luminosities for any tested value of 
$\gamma$\,\footnote{Of course, with the surface boundary condition 
\eqref{Eq:Boundary_surfP} we cannot have super-Eddington surface 
luminosity; thus, we had to extrapolate $\Tb - T_\mathrm{s}$ relations of Sect.\ 
\ref{sec:Env} to higher temperatures to obtain the matching.}. As we do not 
consider mass loss and stellar winds in the current work we had two options: 
change the initial temperature parametrization or explicitly set the initial 
luminosity in the inner envelope. We decided to do the latter. Setting the initial 
luminosity directly requires a separate step in the algorithm to solve for the 
initial temperature given the initial luminosity. We incorporated matching of the 
initial and boundary conditions in this step. In such scheme we have lost direct 
control over the initial temperature, but, as we show in \Sec{sec:Results}, 
direct control over luminosity might be more useful for studying neo-neutron 
stars. Besides, we can still control $T_\mathrm{b,0}$ via $\Tb - 
T_\mathrm{s}$ relations and the fact that $L_\mathrm{s}(T_\mathrm{s}(\Tb)) 
= L_\mathrm{b}$. We can also control $T_{\mathrm{c},0}$ by adjusting the 
initial luminosity (see details in \Sec{sec:Results}). We kept the 
parametrization \eqref{Eq:T_init} to demonstrate how a relatively small change 
in the initial temperature profile can considerably affect the cooling during the 
first $\sim 10^4$ s.



\end{document}